\documentclass[12pt]{article}

\usepackage{amsmath}
\usepackage{amssymb}
\usepackage{epsfig}
\usepackage{graphicx}
\usepackage{cite}
\usepackage{multirow}
\usepackage{longtable}
\usepackage{lscape}
\usepackage{bbm}
\usepackage{color}

\newcommand{\capdef}{}
\newcommand{\mycaption}[2][\capdef]{\renewcommand{\capdef}{#2}%
        \caption[#1]{{\footnotesize #2}}}
\makeatletter
\renewcommand{\fnum@table}{\textbf{\tablename~\thetable}}
\renewcommand{\fnum@figure}{\textbf{\figurename~\thefigure}}
\makeatother

\newcounter{myenumi}

\renewcommand{\themyenumi}{\roman{myenumi}}
{\end{list}}

\newlength{\myem}
\newcounter{mysubequation}[equation]

\makeatletter
\renewcommand{\section}{\@startsection{section}{1}{0em}{-\baselineskip}%
{\baselineskip}{\normalfont\large\bfseries}}
\renewcommand{\subsection}%
{\@startsection{subsection}{2}{0em}{-0.7\baselineskip}%
{0.7\baselineskip}{\normalfont\bfseries}}
\makeatother

\newcommand{\bi}{\begin{itemize}}
\newcommand{\ei}{\end{itemize}}

\newcommand{\be}{\begin{equation}}
\newcommand{\ee}{\end{equation}}
\newcommand{\bea}{\begin{eqnarray}}
\newcommand{\eea}{\end{eqnarray}}

\newcommand{\stheta}{\sin^2 2 \theta_{13}}

\newcommand{\ie}{{\it i.e.}}

\newcommand{\eg}{{\it e.g.}}

\newcommand{\cf}{{\it cf.}}

\newcommand{\etc}{{\it etc.}}
\newcommand{\eq}{Eq.}

\newcommand{\fig}{Fig.}

\newcommand{\Ref}{Ref.}
\newcommand{\Refs}{Refs.}
\newcommand{\Sec}{Sec.}

\newcommand{\App}{App.}

\newcommand{\Tab}{Table}
\newcommand{\Tabs}{Tables}

\newcommand{\equ}[1]{\eq~(\ref{equ:#1})}

\newcommand{\figu}[1]{\fig~\ref{fig:#1}}

\begin{document}

\begin{titlepage}

\renewcommand{\thefootnote}{\alph{footnote}}

\vspace*{-3.cm}
\begin{flushright}
\end{flushright}


\renewcommand{\thefootnote}{\fnsymbol{footnote}}
\setcounter{footnote}{-1}

{\begin{center}
{\large\bf
The Seesaw Mechanism in Quark-Lepton Complementarity} \end{center}}
\renewcommand{\thefootnote}{\alph{footnote}}

\vspace*{.8cm}
\vspace*{.3cm}
{\begin{center} {\large{\sc
 		Florian~Plentinger\footnote[1]{\makebox[1.cm]{Email:}
                florian.plentinger@physik.uni-wuerzburg.de},
 		Gerhart~Seidl\footnote[2]{\makebox[1.cm]{Email:}
                seidl@physik.uni-wuerzburg.de}, and
                Walter~Winter\footnote[3]{\makebox[1.cm]{Email:}
                winter@physik.uni-wuerzburg.de}
                }}
\end{center}}
\vspace*{0cm}
{\it
\begin{center}

       Institut f{\"u}r Theoretische Physik und Astrophysik, Universit{\"a}t W{\"u}rzburg, \\
       D-97074 W{\"u}rzburg, Germany

\end{center}}

\vspace*{1.5cm}

{\Large \bf
\begin{center} Abstract \end{center}  }
We systematically construct realistic mass matrices for
 the type-I seesaw mechanism out of more than 20 trillion possibilities. We use only very generic assumptions from extended
 quark-lepton complementarity, \ie, the leptonic mixing angles between flavor and mass eigenstates
 are either maximal, or parameterized by a single
 small quantity $\epsilon$ that is of the order of the Cabibbo angle
 $\epsilon\simeq\theta_\text{C}$. The small quantity $\epsilon$
 also describes all fermion mass hierarchies. We show that special cases
 often considered in the literature, such as having a symmetric Dirac
 mass matrix or small mixing among charged leptons, constitute only a
 tiny fraction of our possibilities. Moreover, we find that in most cases the spectrum
 of right-handed neutrino masses is only mildly hierarchical. As a
 result, we provide for the charged leptons and neutrinos a selected list of $1 \, 981$ qualitatively
 different Yukawa coupling matrices (or textures) that are parameterized
 by the Cabibbo angle and allow for a perfect fit to current data. In
 addition, we also briefly show how the textures could be generated in explicit models from flavor symmetries. 

\vspace*{.5cm}

\end{titlepage}

\newpage

\renewcommand{\thefootnote}{\arabic{footnote}}
\setcounter{footnote}{0}

\section{Introduction}

The impressive experimental advances that have been made during the past decade in solar~\cite{Fukuda:2002pe,Ahmad:2002ka}, atmospheric~\cite{Fukuda:1998mi}, reactor~\cite{Araki:2004mb,Apollonio:2002gd}, and accelerator~\cite{Aliu:2004sq} neutrino oscillation experiments, have very well
established that neutrinos are massive. Since neutrinos are massless
in the Standard Model (SM), the observation of neutrino masses
provides evidence for new physics, such as an underlying Grand Unified Theory (GUT)~\cite{SU5} (see also
\Ref~\cite{PatiSalam}). It is therefore believed that the smallness of
the absolute neutrino mass scale $m_\nu\simeq 10^{-2}\dots
10^{-1}\:\text{eV}$ compared to the electroweak scale $\sim
10^2\:\text{GeV}$ gives us important information on the nature of the new physics. Today, the most widely
accepted mechanism to generate small neutrino masses is the seesaw mechanism
\cite{typeIseesaw,typeIIseesaw}, in which the smallness of neutrino
masses is linked to the hierarchy between
the electroweak and the GUT scale $M_\text{GUT}\simeq 2\times 10^{16}\:\text{GeV}$~\cite{GUTscale}.

In the type-I seesaw mechanism~\cite{typeIseesaw}, the set of SM
neutrinos $\nu_i$ ($i=1,2,3$ is the generation index) is extended by
three right-handed neutrinos $\nu_i^c$, which are total singlets
under the SM gauge group $G_\text{SM}=SU(3)_c\times SU(2)_L\times
U(1)_Y$. In the basis $(\nu_1,\nu_2,\nu_3,\nu_1^c,\nu_2^c,\nu_3^c)$,
this leads after electroweak symmetry breaking to a complex symmetric
$6\times 6$ matrix
\begin{equation}
 M_\nu=\left(
\begin{matrix}
 0 & M_D\\
 M_D^T & M_R
\end{matrix}
\right),\label{equ:Mnu}
\end{equation}
where 0, $M_D$, and $M_R$ are $3\times 3$ matrices. The upper left
matrix 0 has zero entries since there is no Higgs triplet
that could have directly coupled to the $\nu_i$. The entries in $M_D$
are protected by electroweak gauge invariance and they are therefore
of the order $\sim 10^2\;\text{GeV}$, while the matrix elements of
$M_R$ are of the order of the $B-L$ breaking scale $M_{B-L}\simeq
10^{14}\;\text{GeV}$. After integrating out the right-handed neutrinos, we
arrive at the effective low-energy $3\times 3$ neutrino Majorana mass matrix
\begin{equation}
 M_\text{eff}=-M_DM_R^{-1}M_D^T,\label{equ:Meff}
\end{equation}
which gives rise to neutrino masses of the order $m_\nu\simeq
10^{-2}\;\text{eV}$. The seesaw mechanism is attractive because
$M_{B-L}$ is very close to $M_\text{GUT}$, indicating a GUT-origin
of neutrino masses.

In GUT models, quarks and leptons are unified into multiplets, which
is known as quark-lepton unification and one possibility to explore
GUTs at present energies is to search for signatures of
quark-lepton unification in the fermion mass and mixing
parameters. Most notably, quark-lepton unification has to give an
answer to the question why the quark mixing angles in the Cabibbo-Kobayashi-Maskawa (CKM) matrix $V_\text{CKM}$ \cite{CKM} and the leptonic mixing angles in the Pontecorvo-Maki-Nakagawa-Sakata (PMNS) matrix $U_\text{PMNS}$ \cite{PMNS} are strikingly different. In the quark
sector, all CKM mixing angles are small and can be
approximately written as powers of the Cabibbo angle
$\theta_\text{C}$. In contrast to this, in the lepton sector, only
the reactor angle $\theta_{13}$ is small, whereas the solar angle
$\theta_{12}$ and the atmospheric angle $\theta_{23}$ are both
large. In addition, while the quark and charged lepton mass ratios are
strongly
hierarchical, the neutrino masses exhibit, if any, only a mild hierarchy (for a
recent global fit of neutrino data see, \eg, \Ref~\cite{Schwetz:2006dh}).

Recently, quark-lepton complementarity (QLC)~\cite{qlc} (for an early
approach see \Ref~\cite{Petcov:1993rk}) has been proposed as a
possibility to account for the differences between the quark and
lepton mixings. In QLC,
the quark and lepton mixing angles are connected by the QLC relations
\begin{equation}\label{equ:qlc}
\theta_{12}+\theta_\text{C}\approx\pi/4,\quad\theta_{23}+\theta_{cb}\approx\pi/4, 
\end{equation}
where $\theta_{cb}=\text{arcsin}\:V_{cb}$. The crucial observation is
that sum rules of the types shown in \equ{qlc} can be easily obtained
when the mixing among the neutrinos and among the charged leptons is
described by maximal or CKM-like mixing angles. In this way,
for example, the observed value of the solar angle $\theta_{12}\approx
33^\circ$ could be understood in terms of maximal $(\pi/4)$
and Cabibbo-like ($\theta_\text{C}$) mixing in the individual neutrino
and charged lepton sectors. A complementary approach to the
solar angle seems, on the other hand, to be suggested by the tri-bimaximal mixing scheme~\cite{tribimaximal}. 
The properties of QLC have been studied in various respects: as a result of deviations from bimaximal
mixing~\cite{qlcbimax}, in connection with sum rules~\cite{qlcsumrules}, with emphasis on phenomenological implications~\cite{qlcpheno}, together with
parameterizations of $U_\text{PMNS}$ in terms of $\theta_\text{C}$~\cite{qlcCabibbo},
in view of statistical arguments \cite{Chauhan:2006im}, in
conjunction with renormalization group effects~\cite{qlcRG}, and in model building
realizations~\cite{qlcmodels}.

In \Ref~\cite{Plentinger:2006nb}, we have proposed an {\it extended} QLC, in which the mixing
angles in both the charged lepton and the neutrino sector can take any
value in the sequence $\pi/4,\epsilon,\epsilon^2,\dots$, where
$\epsilon$ is of the order the Cabibbo angle
$\epsilon\simeq\theta_\text{C}$. In this paper, we will implement extended
QLC in the type-I seesaw mechanism by assuming that all the mixing angles
of charged leptons and left- and right-handed neutrinos take their values in this sequence. We also suggest that the mass eigenvalues of $M_D$ and $M_R$ are described by
powers of $\epsilon$ as well. In this approach, the observed large mixing angles $\theta_{12}$ and $\theta_{23}$ can come from
the charged leptons and/or neutrinos.\footnote{For a recent study on the reconstruction of the seesaw mechanism from low-energy data see Ref.~\cite{Casas:2006hf} and large mixing angles coming from the charged lepton sector were also considered, {\it e.g.}, in Ref.~\cite{chargedleptons}.} Moreover, in the neutrino sector, large
mixing angles can originate from $M_D$ and/or $M_R$. We systematically
search for all mass matrices of charged leptons and neutrinos that
satisfy the extended QLC assumptions, and extract all solutions that are consistent
with current data in the CP conserving case.

The paper is organized as follows: We first motivate the assumptions
underlying the extended QLC approach from the phenomenological and model building point of view in \Sec~\ref{sec:motivation}.
This section can be skipped by the reader already familiar with extended QLC. Next, in \Sec~\ref{sec:method}, we describe
the method for constructing all valid charged lepton and seesaw mass
matrices that are compatible with data, demonstrate how to obtain
the corresponding textures, and address further properties of our procedure. For the normal 
neutrino mass hierarchy, we first discuss the full sample of all valid
mass matrices in \Sec~\ref{sec:nhfull},
and then we show a selection of textures in \Sec~\ref{sec:nhtextures}. 
A qualitative discussion of the inverted and degenerate neutrino mass
schemes is included in \Sec~\ref{sec:invdeg} and a summary and
conclusions can be found in \Sec~\ref{sec:summary}. Details of our method
can be found in Appendix \ref{sec:relevant} and \ref{sec:detailsnh}.

\section{Motivation}
\label{sec:motivation}

In this section, we first present a brief review of the observed
hierarchies of fermion masses and mixing angles and relate them to a small
expansion parameter $\epsilon$ that is of the order of the Cabibbo
angle. Then, we discuss two representative $SU(5)$ GUT examples that obtain the observed
mass and mixing parameters from flavor symmetries. The observations
made here will later, in \Sec~\ref{sec:qlc}, serve as a
motivation for the hypotheses in extended QLC.

\subsection{Masses and Mixings of Quarks and Leptons}
\label{sec:masses+mixings}

One of the most striking features of the fermion sector is that the mass and mixing parameters of quarks and charged leptons are
strongly hierarchical. It is well known that these hierarchies can be approximately described
by a small number $\epsilon\simeq 0.2$. In the Wolfenstein
parameterization \cite{Wolfenstein:1983yz}, for example, the CKM matrix
is given by
\begin{equation}
 V_\text{CKM}=\left(
\begin{matrix}
 1-\frac{1}{2}\epsilon^2 & \epsilon & A(\rho-{\text i}\eta)\epsilon^3\\
 -\epsilon & 1-\frac{1}{2}\epsilon^2 & A\epsilon^2\\
A(1-\rho-\text{i}\eta)\epsilon^3 & -A\epsilon^2 &1
\end{matrix}
\right),
\end{equation}
where $\epsilon$ is of the order of the Cabibbo angle
$\theta_\text{C}\simeq 0.2$, and $A,\rho,$ and $\eta$, are order unity
parameters (for an update see Ref.~\cite{Blucher:2005dc}). Order of magnitude wise, the quark mixing angles can be
written in terms of the parameter $\epsilon$ as
\begin{equation}\label{equ:CKMangles}
|V_{us}|\sim\epsilon,\quad |V_{cb}|\sim\epsilon^2,\quad
|V_{ub}|\sim\epsilon^3.
\end{equation}
An interesting connection between the Cabibbo angle and the quark
masses is established by the Gatto-Sartori-Tonin-Oakes relation
$\theta_\text{C}=\sqrt{m_d/m_s}$
\cite{GSTO}, suggesting that
also the fermion mass ratios arise from powers of $\epsilon$. In fact,
the mass ratios of the up and down quarks can,
{\it e.g.}, be crudely represented as powers of $\epsilon$
as\footnote{We are interested here in an $SU(5)$ compatible fit.}
\begin{equation}\label{equ:quarkmassratios}
m_u:m_c:m_t=\epsilon^6:\epsilon^4:1,\quad
m_d:m_s:m_b=\epsilon^4:\epsilon^2:1,
\end{equation}
where $m_b:m_t\sim\epsilon^2$, $m_\tau:m_b\sim 1$, and $m_t\simeq
175\:\text{GeV}$, whereas the mass ratios of the charged leptons are
crudely given by
\begin{equation}
m_e:m_\mu:m_\tau=\epsilon^4:\epsilon^2:1.
\label{equ:clmassratios}
\end{equation}
These mass ratios have all to be understood as order of magnitude
relations and depend on the energy scale. At one loop, in the minimal
supersymmetric standard model (MSSM), the only changes in these relations are $m_c:m_t\sim\epsilon^4$ and
$m_b:m_t\sim\epsilon^3$ at $\sim
10^{15}\;\text{GeV}$~(see, {\it e.g.}, Ref.~\cite{Chankowski:2005qp}). The
changes due to renormalization group (RG) running are thus only comparatively small.

In the neutrino sector, we have a situation that is substantially
different from the charged fermion sectors. Experimentally, the PMNS
matrix reads (\cf~also Ref.~\cite{Gonzalez-Garcia:2004jd})
\begin{equation}
 U_\text{PMNS}=\left(
\begin{matrix}
0.82-0.85 & 0.52-0.57 & \leq 0.12\\
0.26-0.49 & 0.48-0.68 & 0.65-0.76\\
0.27-0.49 & 0.49-0.69 & 0.64-0.75
\end{matrix}
\right),
\end{equation}
which has, unlike $V_\text{CKM}$, large off-diagonal entries. In the standard parameterization, the $1\sigma$ ranges for the
solar and the atmospheric mixing angles are then~\cite{Schwetz:2006dh}
\begin{equation}
 \text{sin}^2\theta_{12}=0.30^{+0.02}_{-0.03},\quad \text{sin}^2\theta_{23}=0.5^{+0.08}_{-0.07},
\end{equation}
whereas we have a $3\sigma$ upper bound on the reactor angle that is
$\text{sin}^2 \theta_{13}\leq 0.041$. The best fit values of the mixing
angles correspond to maximal atmospheric mixing
$\theta_{23}\approx\pi/4$ and large, but not maximal, solar mixing
$\theta_{12}\approx\pi/4-\theta_\text{C}$, and a small reactor angle
$\theta_{13}\lesssim\theta_\text{C}$. The $1\sigma$ bounds on the
solar and atmospheric mass squared differences are \cite{Schwetz:2006dh}
\begin{equation}
\Delta m_\odot^2=(7.9^{+0.3}_{-0.3})\times 10^{-5}\text{eV}^2,\quad
\Delta m_\text{atm}^2=(2.5^{+0.2}_{-0.25})\times 10^{-3}\text{eV}^2.
\end{equation}
The sign of $\Delta m_\odot^2$ is positive, whereas the sign of
$\Delta m_\text{atm}^2$ can be either positive or negative, leading
to currently three possible types of allowed neutrino mass
spectra. Note that we roughly have
\begin{equation}
 \Delta m_\odot^2:\Delta m_\text{atm}^2\sim\epsilon^2.
\end{equation}
It is thus plausible that the neutrino sector, just like the quarks and
charged leptons, is described by the same
control parameter $\epsilon$. The order of magnitude relations for the
neutrino masses may thus be written as
\begin{equation}
 m_1:m_2:m_3=\epsilon^2:\epsilon:1,\quad
 m_1:m_2:m_3=1:1:\epsilon,\quad m_1:m_2:m_3=1:1:1,
\label{equ:numassratios}
\end{equation}
where $m_1,m_2,$ and $m_3$, denote the masses of the 1st, 2nd, and 3rd neutrino
mass eigenstate. The 1st, 2nd, and 3rd equation in
\equ{numassratios} describe a
normal hierarchical (NH), inverse hierarchical (IH), and quasi
degenerate (QD) spectrum, respectively.\footnote{For NH neutrinos, one can compute 
$\epsilon$ from the current best-fit values, which gives $0.15 \lesssim \epsilon
\lesssim 0.22$ ($3\sigma$).}

We thus see that the CKM angles and mass ratios of quarks
and leptons are roughly given by some power $\sim\epsilon^n$ of the
Cabibbo angle $\epsilon\simeq\theta_\text{C}$. Besides that, the
phenomenological QLC relations $\theta_{23}\approx\pi/4-\epsilon^2$ and
$\theta_{12}\approx\pi/4-\epsilon$ involve maximal mixing
angles. In the next section, we will dicuss how these mass and mixing
parameters may be reproduced in models.

\subsection{Examples with Quark-Lepton Unification}
\label{sec:GUTs}
The common appearance of the masses and mixing angles in the quark
and lepton sectors may point towards a quark-lepton unified
theory. One might therefore wonder whether a description of the fermion mass and mixing parameters as given in
Sec.~\ref{sec:masses+mixings} can indeed be obtained in
explicit models. We are interested in models in which the fermion
mixing angles -- prior to going to the mass eigenbasis -- can be maximal
or are given by some power of the Cabibbo angle
$\epsilon\simeq\theta_\text{C}$ which also describes all fermion mass ratios.

For this purpose, let us briefly review two GUT models \cite{Enkhbat:2005xb,Grimus:2002zh} based on $SU(5)$, which
  show that in a quark-lepton unified theory it is (i) actually possible to generate realistic hierarchical fermion mass ratios and
  mixing angles that are described by powers $\sim\epsilon^n$ and that
  (ii) one can predict exact maximal mixing compatible with these
  hierarchies. We view the two GUT models as two representatives of a broad
class of possible realistic models using Abelian (for early work see
Ref.~\cite{earlyU1} and for more recent models see, {\it e.g.}, Refs.~\cite{recentU1,Chankowski:2005qp})
or discrete non-Abelian (for recent studies including the quark sector
  see, {\it e.g.}, Ref.~\cite{discretequarks} and in the context of
  GUTs see, {\it e.g.}, Ref.~\cite{discreteGUTs}) flavor
symmetries.\footnote{For a more complete list of references on discrete
  non-Abelian flavor symmetries see Ref.~\cite{Altarelli:2007cd}.} Our observations can be viewed as a further motivation for
the definition of extended QLC in \Sec~\ref{sec:qlc}, where we will, in
particular, claim that both the neutrinos as well as the charged
leptons can exhibit mixing angles that are maximal and/or $\sim\epsilon^n$. 
In this way, the size of $\theta_{12}$ will then be simply the result
of a QLC-type sum rule.

\begin{figure}[t]
\begin{center}
\includegraphics*[bb = 160 670 420 740]{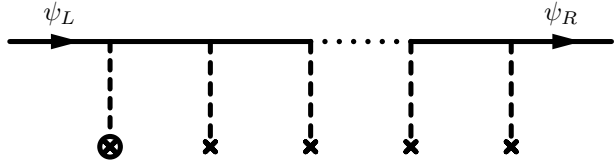}
\caption{\footnotesize Generation of higher-dimension fermion mass terms via the
  Froggatt-Nielsen mechanism. $\psi_L$ and $\psi_R$ are left- and
  right-handed SM fermions. Internal solid lines denote superheavy
  fermions with common mass $M_F$, the circled cross is the usual SM
  Higgs vacuum expectation value (VEV), whereas the
  crosses without circle represent universal VEVs $v$ of SM singlet
  scalars that break the flavor symmetry. After
  integrating out the heavy fermions, the resulting
  mass term is then given by the effective dimension-$n$ operator $\langle
  H\rangle\epsilon^n\overline{\psi}_L\psi_R$, where $\epsilon=v/M_F$
  serves as a small expansion parameter.}
\label{fig:FN}
\end{center}
\end{figure}

{\bf Example 1: Cabibbo-type mass and mixing hierarchies} -- Our first example is a supersymmetric $SU(5)'\times
SU(5)''$ GUT with a $U(1)^N=\Pi_{j=1}^N U(1)_j$ flavor symmetry
group \cite{Enkhbat:2005xb}. It yields, as a result of $U(1)^N$ flavor
symmetry breaking, the masses and mixing angles of quarks and leptons
roughly as powers of $\epsilon$. The $i$th generation $(i=1,2,3)$ is charged
under $SU(5)'\times U(1)_{k_i}$ ($k_i$ labels a suitable
subgroup $U(1)_{k_i}\subset\Pi^N_{j=1}U(1)_j$) in an $SO(10)$ compatible way as ${\bf 10}(-1)_i+{\bf\overline{5}}(3)_i+{\bf 1}(-5)_i$,
where the numbers in parenthesis denote the $U(1)_{k_i}$ charges of the
respective multiplets and the $SU(5)$ singlets are the right-handed
neutrinos. Higgs superfields break the $U(1)^N$ gauge symmetry such that
masses for quarks and leptons arise from higher-dimension terms via
the Froggatt-Nielsen mechanism~\cite{Froggatt:1978nt} illustrated in \figu{FN}.
In the presence of SM singlet scalar ``flavons'', that break the flavor symmetry
by acquiring universal vacuum expectation values (VEVs) $v$ (crosses), and superheavy fermions with
common mass $M_F$, which are charged under the
flavor symmetry (internal solid lines), the mass terms of quarks and leptons become suppressed by integer powers of a small
parameter $\epsilon=v/M_F$ that controls the flavor symmetry
breaking. The integer power of $\epsilon$ is solely determined by the
quantum numbers of the left- and right-handed fermions $\psi_L$ and
$\psi_R$ under the flavor symmetry. As a result, the mass
matrices of the up quarks, down quarks, charged leptons, and neutrinos become
\begin{equation}\label{equ:SU(5)textures}
M_u=m_t\left(
\begin{matrix}
 \epsilon^6 & \epsilon^5 & \epsilon^3\\
 \epsilon^5 & \epsilon^4 & \epsilon^2\\
 \epsilon^3 & \epsilon^2 & 1
\end{matrix}
\right),\quad
M_d=M_\ell^T=m_b\left(
\begin{matrix}
\epsilon^4 & \epsilon^6 & \epsilon^3\\
\epsilon^{10} & \epsilon^2 & \epsilon^2\\
\epsilon^8 & \epsilon^3 & 1
\end{matrix}
\right),\quad
 M_\text{eff}=m_\nu
\left(
\begin{matrix}
\epsilon^4 & \epsilon^2 & \epsilon\\
\epsilon^2 & 1 & 1\\
\epsilon & 1 &1
\end{matrix}
\right).
\end{equation}
The representation of the mass matrices in
Eq.~(\ref{equ:SU(5)textures}) are examples of what we will call in the
following {\bf textures}: These are descriptions of the mass
matrices showing only the order of magnitude (up to order one Yukawa couplings) of the entries in terms
of powers of a small number $\epsilon$ that parameterizes the flavor
symmetry breaking. At an order of magnitude level, the textures in
Eqs.~(\ref{equ:SU(5)textures}) predict for the quarks and charged
leptons the mass ratios and mixing
angles of Eqs.~(\ref{equ:CKMangles}), (\ref{equ:quarkmassratios}), and
(\ref{equ:clmassratios}), whereas the neutrino mass spectrum is of the type $m_1:m_2:m_3=\epsilon:\epsilon:1$. The reactor mixing angle is small and of the order $\theta_{13}\sim
\epsilon$, while the solar and the atmospheric mixing angles are large
and of the orders $\theta_{12}\sim 1$ and $\theta_{23}\sim 1$. A proper choice
of the order one Yukawa couplings then allows to reproduce the solar and atmospheric mixing angles close to the current best fit values.

{\bf Example 2: Maximal mixing} -- Our second example is an
$SU(5)$ GUT with a non-Abelian discrete flavor symmetry between the 2nd
and 3rd generation \cite{Grimus:2002zh}. It predicts a maximal
atmospheric mixing angle $\theta_{23}=\pi/4$ and a zero reactor angle
$\theta_{13}=0$ as a consequence of the discrete symmetry. The basic flavor symmetry of the model is a discrete $Z_2$ exchange symmetry
that implements a maximal atmospheric mixing angle and acts on the $SU(5)$ multiplets of the 2nd and 3rd generation as
$Z_2:\,{\bf \overline 5}_2\leftrightarrow{\bf \overline 5}_3,\:{\bf 10}_2\leftrightarrow{\bf 10}_3,\:{\bf 1}_2\leftrightarrow
{\bf 1}_3,$ where the subscripts denote the generation indices. In addition, the model has a $U(1)$ family number symmetry that does not commute with
the above $Z_2$ generator. As a consequence, the resulting down quark
and charged lepton mass matrices can accommodate the hierarchical down quark and charged lepton
masses and the CKM and PMNS mixing
angles arise in the up quark and the neutrino sector, respectively. The total non-Abelian flavor symmetry enforces in $M_\text{eff}$ a
$\mu-\tau$ exchange symmetry. This predicts a maximal atmospheric mixing angle
$\theta_{23}=\frac{\pi}{4}$ and a vanishing reactor angle
$\theta_{13}=0$. The solar angle $\theta_{12}$, on the other hand, is
large and can easily reproduce the current best fit value.

These two examples show that hierarchical masses and mixings described
by powers of $\epsilon$ as well as maximal mixing can be predicted in
explicit GUT models from flavor symmetries. As a result of this motivation section,
it is therefore plausible to assume that all mixing angles and mass
hierarchies in the quark and lepton sectors are generated by a single
small quantity $\epsilon \simeq \theta_\text{C}$ augmented by possibly
maximal mixing in the up- and/or down-type sectors. This basic
observation will be the basis for our hypotheses underlying extended QLC.

\section{Method}
\label{sec:method}

In this section, we introduce our method for implementing extended QLC in the
seesaw mechanism. For this purpose, we first briefly review the seesaw
mechanism and discuss our notation for
parameterizing the mass and mixing parameters in
\Sec~\ref{sec:massesandmixings}. Next, we define our QLC assumptions
in \Sec~\ref{sec:qlc}, and outline our approach for
generating and selecting the mass matrices of charged leptons and
neutrinos in \Sec~\ref{sec:generating}, \ie, we describe our general procedure. 
While \Sec~\ref{sec:generating} is somewhat qualitative in some points, we 
give more details and a comment on the complexity in Appendix \ref{sec:relevant}. As the next step, we demonstrate how we produce
textures in \Sec~\ref{sec:textureextraction}. Finally, we discuss the role of mass ratios,
input parameters, and RG running in our routine in \Sec~\ref{sec:precision}.

\subsection{Neutrino Mass and Mixing Nomenclature}\label{sec:massesandmixings}
In what follows, we assume that the left-handed SM neutrinos acquire their masses via the type-I seesaw mechanism
\cite{typeIseesaw}.\footnote{The type-II seesaw mechanism \cite{typeIIseesaw}
in extended QLC has already been covered in a previous
analysis of the possible effective $3\times 3$ Majorana neutrino mass matrices
in Ref.~\cite{Plentinger:2006nb} that can be viewed as being generated
by the coupling to some Higgs triplet with small ($\sim 10^{-2}\;\text{eV}$) VEV.} In the type-I seesaw mechanism, the Yukawa
couplings generating the charged lepton and neutrino masses are
\begin{subequations}
\begin{equation}\label{equ:Yukawainteraction}
 \mathcal{L}_\text{Y}=
-(Y_\ell)_{ij}H^\ast\ell_ie^c_j-(Y_D)_{ij}\text{i}\sigma^2H\ell_i\nu_j^c-\frac{1}{2}(M_R)_{ij}\nu^c_i\nu^c_j+\text{h.c.},
\end{equation}
where $\ell_i=(\nu_i,\:e_i)^T$, $e_i^c$, and $\nu^c_i$ ($i=1,2,3$ is the generation
index), are the
left-handed lepton doublets ($\ell_i$), the right-handed charged
leptons ($e^c_i$), and the right-handed SM singlet
neutrinos ($\nu_i^c$). In \equ{Yukawainteraction}, $H$ is the Higgs doublet, $Y_\ell$ and $Y_D$ are
the $3\times 3$ Dirac Yukawa coupling matrices of the charged leptons ($Y_\ell$) and
neutrinos ($Y_D$), $M_R$ is the $3\times 3$ Majorana mass matrix
of the right-handed neutrinos, and $\text{i}\sigma^2$ is the $2\times
2$ antisymmetric tensor. After electroweak symmetry breaking,
$H$ develops a vacuum expectation value
$\langle H\rangle=(0,v/\sqrt{2})^T$, where $v\sim 10^2\:\text{GeV}$,
and the mass terms of the leptons become
\begin{equation}\label{equ:massterms}
 \mathcal{L}_\text{mass}=-(M_\ell)_{ij}e_ie^c_j -(M_D)_{ij}\nu_i\nu_j^c-\frac{1}{2}(M_R)_{ij}\nu^c_i\nu^c_j+\text{h.c.},
\end{equation}
\end{subequations}
where $M_\ell=\langle H\rangle Y_\ell$ is the charged lepton and
$M_D=\langle H\rangle Y_D$ the Dirac neutrino mass matrix. $M_\ell$
and $M_D$ are complex $3\times 3$ matrices that are described by 18
parameters and have entries of the order $\sim 10^2\:\text{GeV}$. The
matrix $M_R$ is complex, symmetric, and described by 12 parameters. It
has matrix elements of the order the $B-L$ breaking scale
$M_{B-L}\sim 10^{15}\:\text{GeV}$. The resulting complex symmetric
$6\times 6$ neutrino mass matrix is given in \equ{Mnu}, and after integrating out the right-handed neutrinos, this gives the effective
$3 \times 3$ neutrino mass matrix $M_\text{eff}$ in \equ{Meff} leading to masses
$\sim 10^{-2}\;\text{eV}$ for the active neutrinos.

To analyze the origin of leptonic mixing, we consider the
diagonalization of the mass terms in $\mathcal{L}_\text{mass}$ in
\equ{massterms} by unitary matrices. Using the convention in \Ref~\cite{Plentinger:2006nb}, we can always write a
general unitary $3\times 3$ matrix $U_\text{unitary}$ as
\begin{subequations}
\begin{equation}\label{equ:unitary}
 U_\text{unitary}=
 \text{diag}\left(e^{\text{i}\varphi_{1}},e^{\text{i}\varphi_{2}},e^{\text{i}\varphi_{3}}\right)\cdot\widehat{U}\cdot\text{diag}\left(e^{\text{i}\alpha_{1}},e^{\text{i}
 \alpha_{2}}, 1 \right)
 ,
\end{equation}
where the phases $\varphi_1$, $\varphi_2$, $\varphi_3$, $\alpha_1$, and $\alpha_2$, take
their values in the interval $\left[0,2\pi\right]$ and
\begin{equation}
 \label{equ:ckm}
 \widehat{U} = \left( 
 \begin{array}{ccc}
   c_{12} c_{13} & s_{12} c_{13} & s_{13} e^{-\text{i}\widehat{\delta}} \\
   -s_{12} c_{23} - c_{12} s_{23} s_{13} e^{\text{i}\widehat{\delta}} &   c_{12} c_{23} -
 s_{12} s_{23} s_{13} e^{\text{i}\widehat{\delta}} & s_{23} c_{13} \\ 
 s_{12} s_{23} - c_{12} c_{23} s_{13}
e^{\text{i}\widehat{\delta}} & -c_{12} s_{23} - s_{12} c_{23} s_{13} e^{\text{i}\widehat{\delta}} & c_{23}
c_{13} 
 \end{array}
 \right) 
\end{equation}
\end{subequations}
is a CKM-like matrix in the standard parameterization with $s_{ij} =
 \sin\hat{\theta}_{ij}$, $c_{ij} = \cos\hat{\theta}_{ij}$, where
$\hat{\theta}_{ij}\in\{\hat{\theta}_{12},\hat{\theta}_{13},\hat{\theta}_{23}\}$ lie all in the first quadrant, {\it i.e.},
$\hat{\theta}_{ij}\in\left[0,\frac{\pi}{2}\right]$, and 
$\widehat{\delta}\in[0,2\pi]$. The matrix $\widehat{U}$ is thus described
 by 3 mixing angles $\theta_{ij}$ and one phase $\delta$, {\it i.e.}, it has 4
 parameters. The matrix $U_\text{unitary}$ has five additional
 phases\footnote{For recent discussions of rephasing invariants in the
 lepton sector see, {\it e.g.}, Ref.~\cite{rephasings}.} and contains therefore in total 9 parameters.

The leptonic Dirac mass matrices $M_\ell$ and $M_D$, and the Majorana
mass matrices $M_R$ and $M_\text{eff}$ are diagonalized by
\begin{equation}
M_\ell=U_\ell M_\ell^\text{diag}U_{\ell'}^\dagger,\quad
M_D = U_D M_D^\text{diag} U_{D'}^\dagger,\quad M_R = U_R
M_R^\text{diag} U_R^T,\quad M_\text{eff}=U_\nu M_\text{eff}^\text{diag}U_\nu^T,
\label{equ:massmatrices}
\end{equation}
where $U_\ell,U_{\ell'},U_D,U_{D'},U_R$, and $U_\nu$, are unitary
mixing matrices,
whereas $M_\ell^\text{diag},M_D^\text{diag},M_R^\text{diag},$ and
$M_\text{eff}^\text{diag}$, are diagonal mass matrices with positive entries. We can
always write the mixing matrices as the products
\begin{equation}\label{equ:mixingmatrices}
 U_x=D_x\widehat{U}_xK_x,
\end{equation}
where $\widehat{U}_x$ are
CKM-like matrices that are parameterized as in
\equ{ckm}, while $D_x$ and $K_x$ are given by
 $D_x=\text{diag}(e^{\text{i}\varphi_1^x},e^{\text{i}\varphi_2^x},e^{\text{i}\varphi_3^x})$
and
$K_x=\text{diag}(e^{\text{i}\alpha_1^x},e^{\text{i}\alpha_2^x},1)$,
where the index $x$ runs over $x=\ell,\ell',D,D',R,\nu$. The phases in $D_x$ and $K_x$ are all in the range
$\varphi_1^x,\varphi_2^x,\varphi_3^x,\alpha_1^x,\alpha_2^x\in[0,2\pi]$.
Each of the matrices $\widehat{U}_x$ in
\equ{mixingmatrices} contains four mixing parameters: three mixing angles
and one phase. We denote the parameters of $\widehat{U}_x$ by
$\theta_{12}^x,\theta_{13}^x,\theta_{23}^x,$ and $\delta^x$. For each
of the matrices $\widehat{U}_x$ in \equ{mixingmatrices}, we define the mixing parameters by
identifying in \equ{ckm} the mixing angles as
$\hat{\theta}_{ij}\rightarrow \theta_{ij}^x$, and the phase as
$\widehat{\delta}\rightarrow\delta^x$.

The PMNS matrix is given by
\begin{equation}\label{equ:pmnspara}
U_\text{PMNS}=U_\ell^\dagger U_\nu=\widehat{U}_\text{PMNS}K_\text{Maj},
\end{equation}
where $\widehat{U}$ is a CKM-like matrix parameterized as in
 \equ{ckm}, and
 $K_\text{Maj}=\text{diag}(e^{\text{i}\phi_1},e^{\text{i}\phi_2},1)$
 contains the Majorana phases $\phi_{1}$ and $\phi_2$. The CKM-like matrix $\widehat{U}_\text{PMNS}$ in \equ{pmnspara} is
described by the solar angle $\theta_{12}$, the reactor angle
$\theta_{13}$, the atmospheric angle $\theta_{23}$, and the Dirac
CP-phase $\delta$, which we identify in the standard
parameterization of \equ{ckm} as
$\hat{\theta}_{ij}\rightarrow\theta_{ij}$ and
$\widehat{\delta}\rightarrow\delta$. The PMNS matrix has thus 3 mixing
angles and 3 phases and contains therefore 6 physical
parameters.

Let us next express $M_\text{eff}$ in terms of the mass eigenvalues
and mixing angles introduced above. Inserting \equ{mixingmatrices}
into \equ{massmatrices}, we find
\begin{subequations}\label{equ:mdmr}
\begin{eqnarray}
M_D&=&D_D \widehat{U}_D K_D M_D^{\mathrm{diag}} K_{D'}^*
\widehat{U}_{D'}^\dagger D_{D'}^*\,,\\
M_R^{-1}&=&D_R^*\widehat{U}_R^\ast K_R^*
(M_R^{\mathrm{diag}})^{-1} K_R^* \widehat{U}_R^\dagger D_R^*~.
\end{eqnarray}
\end{subequations}
The effective neutrino mass matrix $M_\text{eff}$ in \equ{Meff} can
thus be written as
\begin{subequations}\label{equ:mthmexpeff}
\begin{equation}
M^\text{th}_\text{eff}=-D_D
\widehat{U}_D\tilde{K}M_D^{\mathrm{diag}}\widehat{U}_{D'}^\dagger
\tilde{D}
\widehat{U}_R^* (K_R^*)^2
(M_R^{\mathrm{diag}})^{-1}\widehat{U}_R^\dagger
\tilde{D}\widehat{U}_{D'}^*
M_D^{\mathrm{diag}}\tilde{K}\widehat{U}_D^T D_D~,
\label{equ:meffth}
\end{equation}
where we have introduced $\tilde{K}=K_D^\ast K_{D'}$ and
$\tilde{D}=D_{D'}^\ast D_R^\ast$. We have denoted the
parameterization of $M_\text{eff}$ in \equ{meffth} by an extra
superscript ``th'' for ``theoretical'', since none of the mass and mixing
parameters on the right-hand side of \equ{meffth} are directly
measurable in neutrino oscillations. Note
that in the CP conserving case, the matrix $(K_R^\ast)^2$ drops out of
the expression for $M_\text{eff}$ in \equ{meffth}. Equivalently to
\equ{meffth}, using Eqs.~(\ref{equ:pmnspara}) and
(\ref{equ:mixingmatrices}) in the expression for $M_\text{eff}$ in
\equ{massmatrices}, we can write $M_\text{eff}$ also in the
parameterization
\begin{equation}
M_\text{eff}^\text{exp}=D_\ell\widehat{U}_\ell K_\ell\widehat{U}_\text{PMNS}K^2_\text{Maj}
M^\text{diag}_\text{eff}\widehat{U}^T_\text{PMNS}K_\ell\widehat{U}_\ell^TD_\ell~,
\label{equ:meffexp}
\end{equation}
\end{subequations}
where we have chosen the superscript ``exp'' for ``experimental'', to label the
representation of $M_\text{eff}$ in \equ{meffexp}, since
$M_\text{eff}^\text{exp}$ involves the matrices $M_\text{eff}^\text{diag}$ and
$U_\text{PMNS}$ containing the experimentally accessible mass and
mixing parameters. Note that in the CP conserving case,
$(K_\text{Maj})^2$ drops out of the expression for
$M_\text{eff}^\text{exp}$. It is clear that $M_\text{eff}^\text{th}=M_\text{eff}^\text{exp}=M_\text{eff}$, since
$M_\text{eff}^\text{th}$ and $M_\text{eff}^\text{exp}$ are just different
parameterizations of $M_\text{eff}$.

\subsection{Hypotheses for Extended QLC}
\label{sec:qlc}

We will now formulate the assumptions underlying extended QLC in the
 type-I seesaw mechanism. Motivated by the discussion of the GUT examples in
 \Sec~\ref{sec:GUTs}, extended QLC will in this context include
 assumptions on (i) the mixing parameters of $U_\ell,U_D,U_{D'},$
and $U_R$, and assumptions on (ii) the eigenvalues of the mass
matrices $M_\ell,M_\text{eff},M_D,$ and $M_R$.

{\bf Mixing angles} -- Consider first the possible mixing angles in extended QLC. Following the idea of
\Ref~\cite{Plentinger:2006nb}, we begin by assuming that all mixing
angles parameterizing in \equ{massmatrices} the mixing matrices $U_\ell,U_D,U_{D'}$, and $U_R$, can
a priori take any of the values in the sequence
$\pi/4,\epsilon,\epsilon^2,\dots$, where $\epsilon$ is a small
number. Motivated by the quark sector, we will take $\epsilon\simeq
0.2$, {\it i.e.}, we assume that $\epsilon$ is of the order the Cabibbo angle.
This applies the concept of
extended quark-lepton complementarity in \Ref~\cite{Plentinger:2006nb}
to the mixing matrices $U_D$, $U_{D'},$ and $U_R$, that diagonalize the
renormalizable neutrino mass terms in \equ{massterms}.
Since the current $1\sigma$ error on the leptonic mixing angles is at most
of the order $\epsilon^2$ (see, \eg, \Ref~\cite{Schwetz:2006dh}), we
will truncate the sequence of mixing angles
$\pi/4,\epsilon,\epsilon^2,\dots$ after the element $\epsilon^2$ and
identify there all terms of the order $\epsilon^n$
with $n\geq 3$ simply by ``$0$''. In other words, in extended QLC, we restrict the possible range of mixing
angles $\theta_{ij}^x$ ($x=\ell,D,D',R$) to the set of values
$\theta_{ij}^x\in\{\pi/4,\epsilon,\epsilon^2,0\}$, where ``$0$'' represents
mixing angles $\sim\epsilon^n$ with $n\geq 3$. Since we want to
compare with current neutrino data, the assumptions on the mixing angles are
formulated at low energies $\sim 1\;\text{GeV}$. The impact of RG running when assuming these
angles at a high scale will be discussed later in \Sec~\ref{sec:precision}.
Note that we cannot just rotate away $U_\ell$, because this would make
our mixing angle assumptions as powers of $\epsilon$ meaningless if induced by
an underlying theory, such as a flavor symmetry.  

{\bf Mass eigenvalues} -- Next, let us specify the types of mass eigenvalues that we assume in
extended QLC. We have to distinguish two types of mass eigenvalues --
those of the lepton mass matrices $M_\ell$ and $M_\text{eff}$ in the
low-energy effective theory, and those of the Dirac and heavy Majorana mass
matrices $M_D$ and $M_R$ of the neutrinos.

We assume that the mass spectra in the low energy effective theory are those of
\equ{clmassratios} and \equ{numassratios}. That means we assume
$m_e:m_\mu:m_\tau=\epsilon^4:\epsilon^2:1$ for the charged leptons,
whereas we have in the neutrino sector
$m_1:m_2:m_3=\epsilon^2:\epsilon:1$ for a NH,
$m_1:m_2:m_3=1:1:\epsilon$ for an IH, and
$m_1:m_2:m_3=1:1:1$ for a QD neutrino mass spectrum. As we will
 discuss later, our results are actually completely independent 
from the details of the charged lepton spectrum.

Let us now consider the mass eigenvalues of the Dirac and Majorana
mass matrices $M_D$ and $M_R$. We denote the mass
eigenvalues of $M_D$ by $m_1^D,m_2^D,$ and $m_3^D$, and
the mass eigenvalues of $M_R$ by $m_1^R,m_2^R,$ and $m_3^R$. Notice that these are not directly
observable at low energies, but as before, we write the mass eigenvalues
of $M_D$ and $M_R$ as powers of $\epsilon$. Similar to
Eqs.~(\ref{equ:clmassratios}) and (\ref{equ:numassratios}), we will therefore
parameterize the mass eigenvalues $M_D$ and $M_R$ as
\begin{equation}
m_1^D:m_2^D:m_3^D=\epsilon^a:\epsilon^b:\epsilon^c\quad\text{and}\quad
m_1^R:m_2^R:m_3^R=\epsilon^{a'}:\epsilon^{b'}:\epsilon^{c'},
\label{equ:mdmrdiag}
\end{equation}
where $a,b,c,a',b',$ and $c'$, are suitable non-negative integers
$\leq 2$, and we define the absolute mass scales by
$m_3^D=m_D\epsilon^c$ and $m_3^R=M_{B-L}\epsilon^{c'}$. As we will
discuss in Appendix \ref{sec:paramhier}, one can restrict the
possible range of the integers in \equ{mdmrdiag} to a fairly small set
of numbers, such that it is sufficient to test all
possible combinations up to second order in $\epsilon$.

In total, we see that in our hypotheses for extended
QLC all mass hierarchies and small mixing angles
become described by powers of the hierarchy parameter $\epsilon$. In
\Sec~\ref{sec:precision}, we will discuss the validity and precision
of our assumptions in extended QLC when taking, {\it e.g.},
GUT relations among fermion masses and RG effects
into account.

\subsection{Procedure: Systematic Construction of the Parameter Space}
\label{sec:generating}

\begin{figure}[htb]
\hspace*{3cm}\includegraphics[width=13cm]{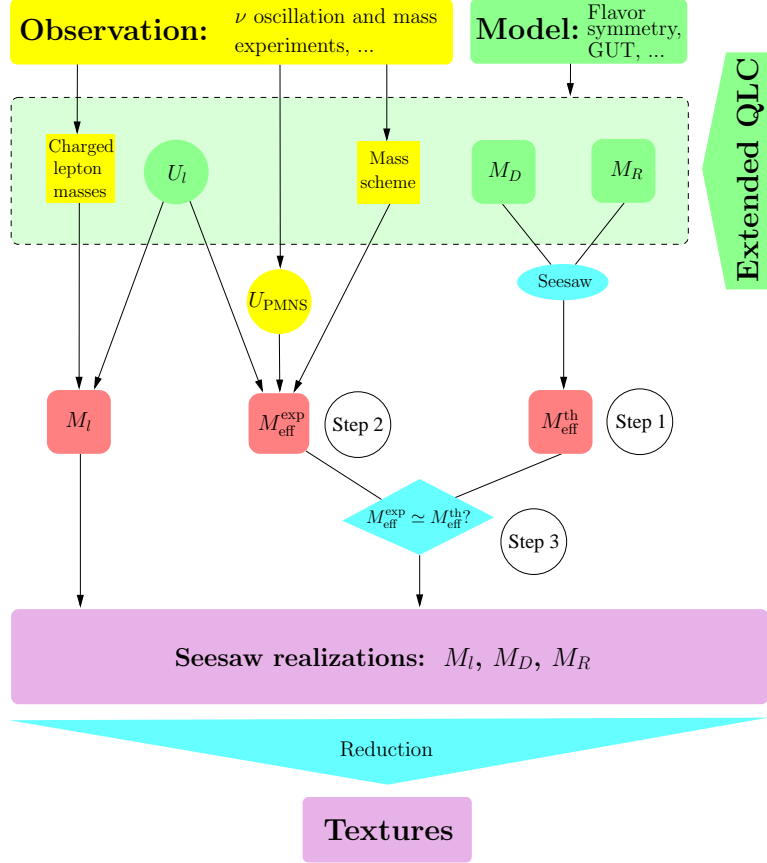}
\caption{\footnotesize Procedure for obtaining the seesaw realizations and
 texture sets in extended QLC.\label{fig:structure}
}
\end{figure}

Let us now describe our three-step procedure for
generating all textures $M_\ell$, $M_D$, and $M_R$, which
satisfy extended QLC for the seesaw mechanism (\cf, \figu{structure} for illustration):

{\bf First step} -- We generate all possibilities for the
 effective neutrino mass matrix $M^\text{th}_\text{eff}$ in \equ{meffth}. 
Here, we assume that the mixing angles entering
$U_D,U_{D'},$ and $U_R$, can take any values in the set
\begin{equation}
 \theta_{ij}^x\in\{\pi/4,\epsilon,\epsilon^2,0\},
\label{equ:thetaxij}
\end{equation}
where $x=D,D',R$. Moreover, we suppose in
\equ{meffth} that $M^\text{diag}_D$ and $M^\text{diag}_R$ are on the general forms as
given in \equ{mdmrdiag} with eigenvalues $ 1, \epsilon$ or$ \epsilon^2$. For
simplicity, we will confine ourselves to the CP conserving
case\footnote{Note that in the case of CP violation, some textures may
  change due to cancellations, so the number of textures will
  increase, like one would expect. Nevertheless, a complete systematic
  analysis (all phases between 0 and $2\pi$ are allowed) is up to now
  not possible because of lack of computing power. This may change in some years.} were
all phases are taken from the set
$\delta^x,\varphi_1^x,\varphi_2^x,\varphi_3^x,\alpha_1^x,\alpha_2^x\in\{0,\pi\}$.

{\bf Second step} -- We generate all possibilities for the neutrino mass matrix $M^\text{exp}_\text{eff}$ in \equ{meffexp}. 
For $U_{\text{PMNS}}$, we use values motivated by the current best-fit values. In particular,
we use the following input, which could be experimentally confirmed or
rejected in the coming years\footnote{See, \eg,
  \Refs~\cite{Huber:2004ug,Antusch:2004yx} for long-baseline experiments on a scale of the coming ten years,
  \Refs~\cite{Minakata:2004jt,Bandyopadhyay:2004cp} for an up scale
  reactor $\theta_{12}$ measurement, \Ref~\cite{Barger:2006kp} for the
  potential of various different superbeam upgrades, and
  \Ref~\cite{nufact} for a  neutrino factory measurement.}
\begin{equation}
\theta_{12} = \pi/4 - \epsilon, \quad \theta_{13} = 0, \quad \theta_{23}=\pi/4 \, 
\label{equ:bfinput}
\end{equation}
with $\epsilon=0.2$. These values represent the current best-fit values~\cite{Schwetz:2006dh}
very closely.\footnote{Note that these values correspond not only
  to the best-fit values, but are also often considered as an
  interesting symmetry limit in ``exceptional'' \cite{Altarelli:2007cd} neutrino mass models.} A different choice for these parameters can be equally well applied, but it will change the final results.

For $U_\ell$, we follow \equ{thetaxij}, and for the phases, we test all real possibilities.
Furthermore, we insert  the neutrino
mass spectra given in \equ{numassratios} into $M_\text{eff}^\text{diag}$, and again test all possibilities.

{\bf Third step} -- Next, we match all possibilities from step~1 and step~2, \ie,
we select all parameter combinations for which
\begin{equation}
 M_\text{eff}^\text{th}|_{\epsilon=0.2} \simeq M_\text{eff}^\text{exp}|_{\epsilon=0.2} \, 
\label{equ:matching}
\end{equation}
at $\mathcal{O}(\epsilon^3)$. For details (and the exact numerical
implementation) of the matching procedure, see Appendix \ref{sec:comparison}; in particular, note that our procedure automatically factors out the
overall neutrino mass scale, \ie, $m_\nu = m_D^2/M_R$ will be automatically satisfied. In the following, 
we will call a (seesaw) {\bf realization} a valid set 
$\{M_D,M_R,U_\ell\}$, or more precisely, a combination of all involved mixing parameters, phases,
and mass hierarchies, for which \equ{matching} is fulfilled. In other
words, a realization is
thus a set of input parameters compatible with current experimental best-fit values which describes $M_\text{eff}^\text{th}$ and the $6\times 6$ matrix $M_\nu$ completely, and it contains the 
left-handed charged lepton mixing. In total, our procedure requires that we systematically scan 20 trillion different possible realizations. For details on the complexity, see Appendix \ref{sec:counting}.

Note that in our procedure the observed large leptonic mixing angles can
be generated either in the charged lepton sector and/or the neutrino
sector. Furthermore, in the neutrino sector, large mixings can arise from the
Dirac neutrino and/or the Majorana mass matrix of the right-handed neutrinos.
This means that we not make any special assumptions simplifying the
structure of the seesaw mechanism, such as taking $M_\ell$ or $M_R$ to be diagonal, or
$M_D$ symmetric.

\subsection{Texture Extraction and Order Unity Couplings}
\label{sec:textureextraction}

Let us now describe how we extract the textures for the charged leptons
 and neutrinos from the seesaw realizations that satisfy
 \equ{matching}. In the course of applying the three step procedure described in
 Sec.~\ref{sec:generating}, we have already produced for each
 realization the pair of matrices $M_D$ and $M_R$. In the same way, we determine
 for each valid realization the charged lepton mass matrix\footnote{Note that we choose, for simplicity, the right-handed charged
 lepton mixing matrix to be the unit matrix
 $U_{\ell'}=\mathbbm{1}$. This choice is, however, not limiting our
 procedure since $U_{\ell'}$ does not enter into $U_\text{PMNS}$.} $M_\ell$ by
 rotating to the left-handed flavor basis as
 $M_\ell=U_\ell M_\ell^\text{diag}$, where $M_\ell^\text{diag}$
 contains the masses given in \equ{clmassratios}.
Next, we analytically expand the mass matrices in $\epsilon$ as
\begin{equation}
 M_x=M_x^{(0)}+M_x^{(1)}\epsilon+M_x^{(2)}\epsilon^2+\mathcal{O}(\epsilon^3),
\label{equ:expand}
\end{equation}
where $x=\ell,D,R,$ and identify for each matrix element the leading
contribution as the lowest order in $\epsilon$. The
texture for $M_x$ is
then found by substituting each matrix element of $M_x$ by its leading power $\epsilon^k$.
For $k \ge 3$, we take $\epsilon^k \rightarrow 0$.
In doing so, we always ``drop'' the order one coefficients that multiply the
leading powers $\epsilon^k$ and round them to one (unless they are
zero).  For a given realization, we call the collection of textures for $M_\ell,M_D,$ and $M_R$ a {\bf texture
  set}. For any texture set obtained in this way, we argue that one can always
adjust the order unity coefficients such that the Yukawa couplings are
brought in perfect agreement with data. An important prerequisite for
this statement is that the different orders in the expansion in
\equ{expand} do not ``interfere'' with each other, \ie, the involved
coefficients are really of order unity. This property for the
coefficients will be checked below. It is important to keep in mind that, in general,
more than one realization may lead to the same texture set. Because of
the reduction from 
in general several realizations to one texture set, we will call our
texture set producing technique ``texture reduction''. Note also the
important fact that the texture reduction is based on an
{\it analytic expansion} in $\epsilon$ as opposed to just using powers of
$0.2$ for a purely numerical fit of the mass matrix elements.

As an example for texture reduction, consider the following mass matrix $M_D$ (\cf, texture/realization \#1 in  \Tab~\ref{tab:seesawtextures}):
\begin{equation*}
\label{equ:textureexample}
M_D=m_D\,\left(
\begin{array}{ccc}
-\frac{\epsilon^2}{\sqrt{2}} & -\epsilon & 0 \\
-\epsilon^2 & -\frac{1}{\sqrt{2}}+\frac{\epsilon^2}{2\sqrt{2}} & \frac{1}{\sqrt{2}}-\frac{\epsilon^2}{2\sqrt{2}} \\
 \frac{\epsilon^2}{\sqrt{2}} & -\frac{\epsilon^2}{\sqrt{2}} & -\epsilon+\frac{\epsilon^2}{\sqrt{2}}
\end{array}
\right)\longrightarrow\left(
\begin{array}{ccc}
\epsilon^2 & \epsilon & 0 \\
\epsilon^2 & 1 & 1 \\
\epsilon^2 & \epsilon^2 & \epsilon
\end{array}
\right).
\end{equation*}
Here, ``$\rightarrow$'' symbolizes, up to an overall mass scale,
 the identification of the leading order terms in the expansion in
 $\epsilon$ that contribute to the mass matrix elements in
 $M_D$. The matrix on the right hand side of $\rightarrow$ then
 represents the texture corresponding to the mass matrix $M_D$ on the left hand side.

\begin{figure}[t]
\begin{center}
\includegraphics[width=0.3\textwidth]{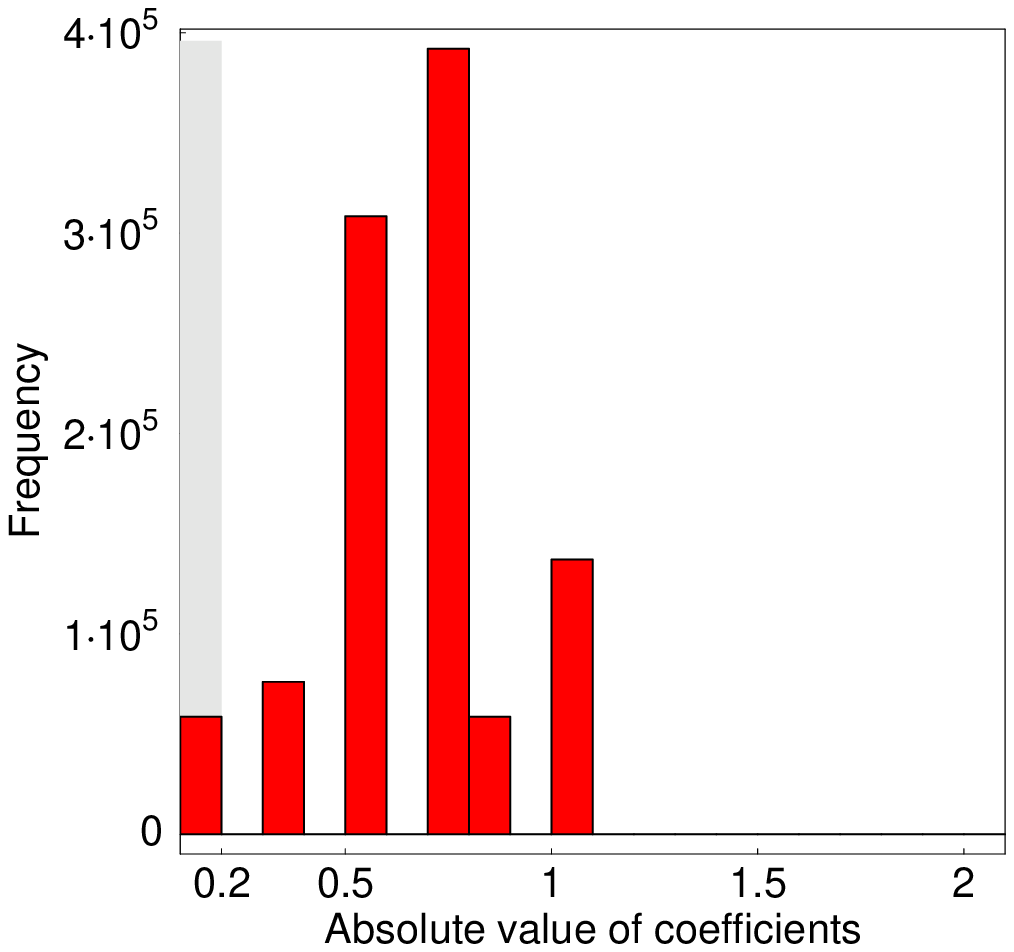} \hspace*{0.03\textwidth}
\includegraphics[width=0.3\textwidth]{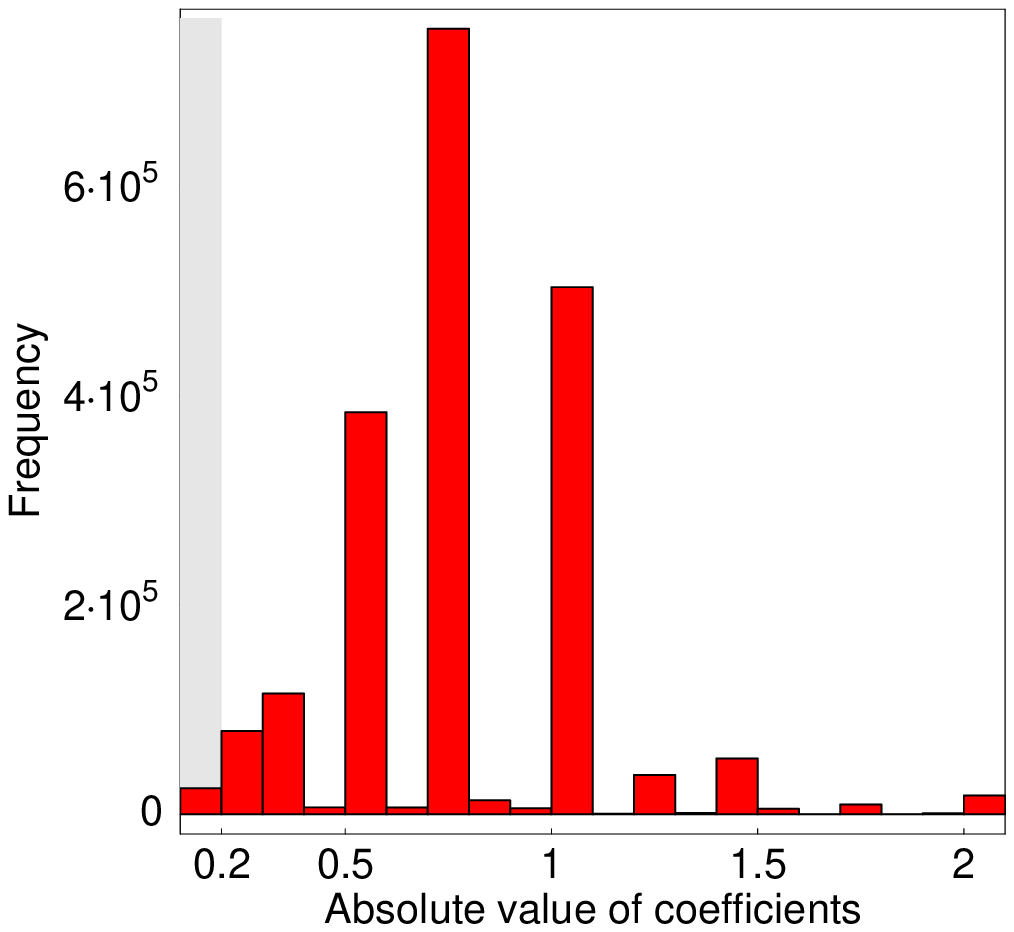} \hspace*{0.03\textwidth}
\includegraphics[width=0.3\textwidth]{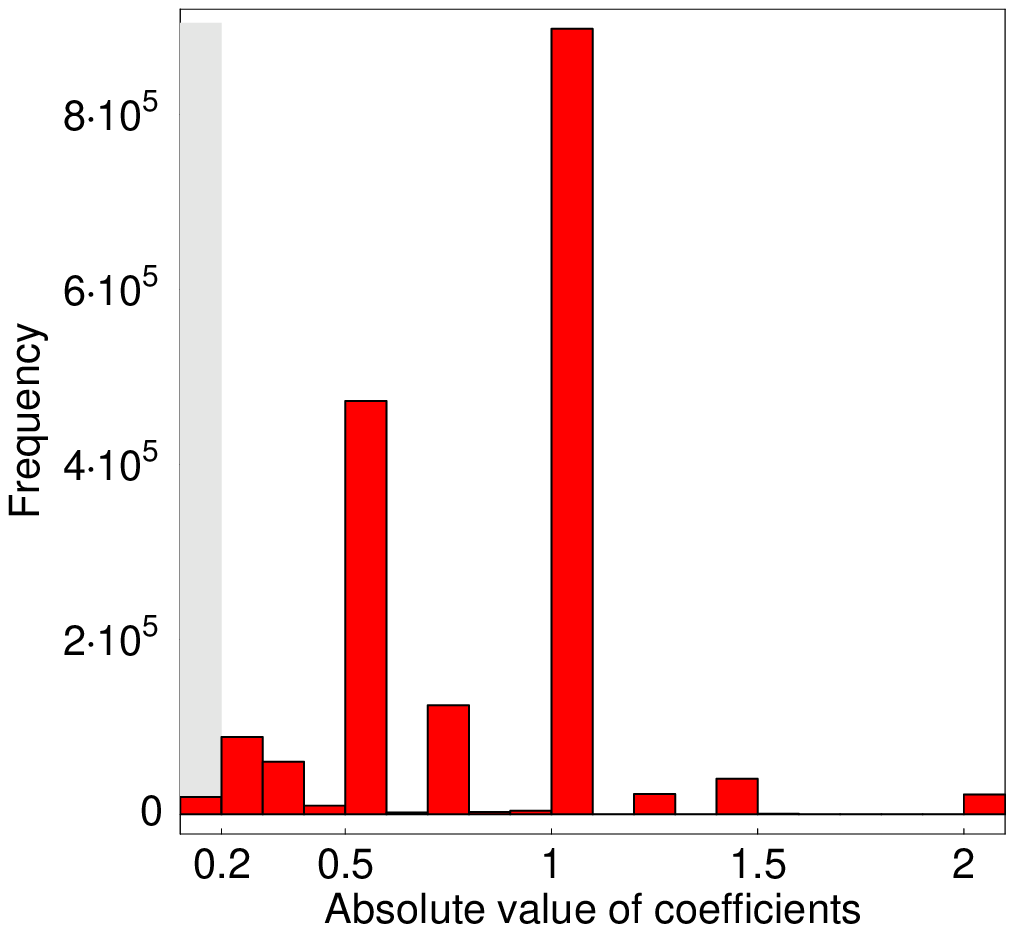}
\end{center}
\caption{\label{fig:coeff} \footnotesize Distribution of the order one
  couplings in $M_\ell$ (left), $M_D$ (center) and $M_R$ (right) for
  all valid seesaw realizations of NH neutrinos (of all orders and for all
  matrix elements). The gray-shaded region marks the area in which the
  order one couplings are smaller than $\epsilon$.}
\end{figure}

In order for the expansion in \equ{expand} to be useful, the Yukawa
couplings should be of order unity
(at least if one wants not to rely on some sort of fine-tuning). The
expansion suggests a criterion for what ``order unity'' actually
means: the coefficients should lie in the interval
$[\epsilon,\dots,\epsilon^{-1}]\simeq[0.2,\dots,5]$, such that they
would not imitate different orders in $\epsilon$. In \figu{coeff}, we
show to which extent this condition
is indeed satisfied for the valid realizations. \figu{coeff} depicts the
coefficients of the $M_x^{(i)}$ in \equ{expand} for $M_\ell$ (left), $M_D$
(center) and $M_R$ (right) for all valid seesaw realizations and NH
neutrinos (including all orders and for all matrix elements). Interestingly, it
turns out that in 99.9\% of all cases, the order unity Yukawa
couplings lie in the range $\epsilon,\dots,\epsilon^{-1}$, which
justifies the expansion of the textures in $\epsilon$. For example, for $M_\ell$
and $M_D$, the peaks at around $0.7 \simeq 1/\sqrt{2}$ are predominant.
Note that, though in \figu{coeff} the first bin corresponds to coefficients smaller
than $\epsilon$, our mapping of textures is unambiguous. This is
because in the cases where the leading order coefficient of a matrix
element becomes very small ($<0.2$), the other coefficients become
also very small -- with the exception of $0.10\%$ of all cases for the
matrix elements.\footnote{All these exceptions appear only in $M_D$.} As a
consequence,  the leading order term indeed rarely numerically interferes
with the higher orders. This picture would change if $\epsilon$ was
larger. For instance, for $\epsilon=0.5$, the leading order
identification for $M_D$ fails in $0.43\%$ of all cases, and for $M_R$
it fails in $0.01\%$ of all cases, which is, however, still only a
remarkably small fraction of all cases.

\subsection{Dependence on Input and Renormalization Group Effects}
\label{sec:precision}

Let us now discuss how the choice of the input parameters from
measurement affects our results. First of all, we use the current
best-fit values as an input (\cf, \equ{bfinput}). Any other choice 
of experimental input parameters will give different results, but
our procedure can be equally well applied to any other preferred choice of input
values. We have chosen these input values because these
results can be expected to remain valid for the next ten years or so,
unless the best-fit values change (such as if $\stheta \sim 0.1$
were indeed found). We have also computed the dataset for different
values of $\theta_{13}$, but a presentation of these results
would clearly exceed the scope of this paper.

As far as the charged lepton mass spectra in \equ{clmassratios} are concerned, a variation of the mass ratios will not have any effect at all
on the selection and extraction of the texture sets: Any change in the
spectrum drops completely out of our routine. The particular choice
of the charged lepton mass hierarchies in \equ{clmassratios} has been
motivated by comparison with the down quark spectrum in
$SU(5)$. Choosing a different parameterization of the charged lepton masses, for example with different powers of $\epsilon$, will of
course have an effect on the form of the extracted texture sets for
$M_\ell$, but it does not affect the selection of the realizations in any way. In particular, a modification of the charged
lepton mass spectra in \equ{clmassratios}, {\it e.g.}, to implement the
Georgi-Jarlskog relation $m_\mu:m_\tau=3\,m_s:m_b$
\cite{Georgi:1979df}, would not change any of our results for the
textures. In fact, one can easily obtain the textures of the charged leptons for any other choice of the hierarchy by using $U_\ell$ directly.

The extended QLC hypotheses in \Sec~\ref{sec:qlc}
should hold at high energies such as $M_\text{GUT}$, but they are
compared in our method with low-energy data. We therefore have
to address the stability of the extended QLC assumptions under RG running from the high scale down to, say, around
$\sim 1 \, \text{GeV}$. Generally, the empirical QLC sum rule
$\theta_{12}=\pi/4-\theta_\text{C}$ is satisfied up to a
precision of about $\sim 1^\circ$. It is
therefore reasonable to take in our routine nonzero mixing angles
$\theta^x_{ij}$ into account that can be as small as
$\epsilon^2\sim 2^\circ$. Moreover, it is known that the
Cabibbo angle $\theta_\text{C}$ does practically not run~\cite{Arason:1991hu},
and $V_{cb}\sim \epsilon^2$ changes
typically only by a factor smaller than 2 when running from $\sim
1\,\text{GeV}$ up to the Planck scale $\sim 10^{19}\,\text{GeV}$ \cite{Arason:1992eb}.

Let us have a more precise look at the RG evolution
of neutrino masses and mixings~\cite{neutrinoRGs}. First, note that,
due to the smallness of the charged lepton Yukawa couplings, the
running of a possibly maximal atmospheric mixing angle $\theta_{23}$ is negligible, unless
one works in the MSSM with large $\text{tan}\,\beta$~\cite{Antusch:2004yx}. Simple
expressions for the running of lepton mixing angles have been recently
presented in \Ref~\cite{Dighe:2007ks}: When running $M_\text{eff}$ from the GUT scale
down to low energies, the corrections to the leptonic mixing angle
$\theta_{ij}$ are smaller than $\sim|m_i+m_j|^2(|m_i|^2-|m_j|^2)^{-1}\times 10^{-2}$, where
$m_i$ and $m_j$ are the eigenvalues of the $i$th and $j$th neutrino
mass eigenstates of $M_\text{eff}$ at the GUT scale. An appreciable
running of leptonic mixing angles can thus only be expected in the IH
or QD case. For NH neutrinos, however, the corrections are $\lesssim 1^\circ$ and, thus,
negligible. Moreover, a tuning of phases always allows to switch off
completely any RG effect on neutrino mixing angles -- even in the case
of inverse hierarchical and degenerate
neutrinos~\cite{Dighe:2007ks}. A similar result has been obtained in
the bottom-up approach in Ref.~\cite{Ellis:2005dr}, where the starting
point are the fixed low-energy observables. In addition, while the overall neutrino mass scale is affected by RG running, the neutrino mass ratios are hardly changed. Since our results should be very stable under RG running for the NH case (irrespective of the
phases), we will focus in this paper on this type of hierarchy.

\section{Currently Allowed Realizations for the NH Case} 
\label{sec:nhfull}

In this section, we focus on the constructed set of currently allowed
seesaw realizations for a normal neutrino mass hierarchy (NH case). We
obtain $173 \, 084$ different realizations for the NH case, which
reduce to $8 \, 030$ cases if one does not count different phase combinations
as different cases. This leads to $1 \, 981$ different texture sets, \ie, different combinations
of $M_\ell$, $M_D$, and $M_R$. Naturally, we cannot show all of these possibilities
in this paper. In \Sec~\ref{sec:nhtextures}, we will therfore apply some selection criteria to reduce this dataset further and present the texture sets which seem to be most interesting to us.
In this section, however, we discuss general features and some
statistics of the constructed seesaw realizations for NH neutrinos.

We concentrate on NH also because RG effects on neutrino mass ratios and mixing
angles are expected to be small in this case (see
\Sec~\ref{sec:precision}). This means that our generic assumptions,
which may hold at some high energy scale (say at $\sim 10^{16}\;\text{GeV}$), do not change significantly when
running down to low energies where we match to experiment.
In the NH case, one can easily diagonalize $M_{\text{eff}}$ for the allowed realizations in order to check
that \equ{matching} produces observables in agreement with current data. We have done this exercise:
We have determined $U_\nu$ by diagonalizing $M_{\text{eff}}$ and
computed $U_{\mathrm{PMNS}} = U_\ell^\dagger U_\nu$ employing the
corresponding matrix $U_\ell$. From $U_{\mathrm{PMNS}}$, one can read off the
mixing angles just as described in \Ref~\cite{Plentinger:2006nb}.\footnote{See,
  {\it e.g.}, also \Ref~\cite{King:2003jb} for a discussion of
  neutrino mass matrix diagonalization.}
Note that we do not expect to reproduce exactly our input values in
\equ{bfinput}, since we do not require exact matching precision in
\equ{matching} (see also Appendix \ref{sec:comparison}).

\begin{figure}[t]
\begin{center}
\includegraphics[width=0.5\textwidth]{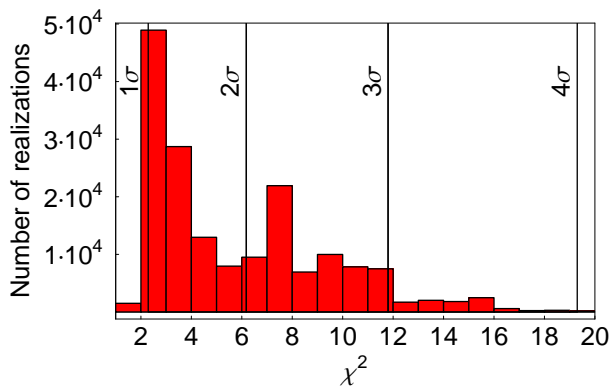}
\end{center}
\caption{\label{fig:allchi2}\footnotesize Distribution of valid seesaw
  realizations for the NH case as
a function of $\chi^2$ as defined in \equ{chi2}. The values of $\theta_{13}$
are all in agreement with current data.}
\end{figure}

To describe the compatibility of a realization with current data, we
use the performance indicator
\begin{equation}
\chi^2 \equiv \left( \frac{\sin^2 \theta_{12}  - 0.3}{0.3 \times \sigma_{12}} \right)^2 + \left( \frac{\sin^2 \theta_{23} -  0.5}{0.5 \times \sigma_{23}} \right )^2~,
\label{equ:chi2}
\end{equation}
(\eg, $\chi^2=11.83$ corresponds to a $3\sigma$ CL exclusion for 2
d.o.f.). This corresponds to a Gaussian $\chi^2$ approximation in
$\sin^2 \theta_{12}$ and $\sin^2 \theta_{23}$ with the current
best-fit values. For the relative $1\sigma$ errors, we use
$\sigma_{12} \simeq 9\%$ (for $\sin^2\theta_{12}$) and $\sigma_{23}
\simeq 16\%$ (for $\sin^2\theta_{23}$)~\cite{Schwetz:2006dh}.
Note that we only find $\sin^2 \theta_{13} \ll 0.04$ below the current
bound, \ie, we do not have to impose an additional selection criterion.
\figu{allchi2} shows the distribution of the valid seesaw realizations as
a function of $\chi^2$ defined in \equ{chi2}. Obviously,
\equ{matching} already ensures that the neutrino mixing angles of
each realization are compatible with current bounds. It turns out that
in all valid cases $\theta_{13}\ll 1^\circ$ and only 6.5\% of the realizations lead to
$11.83\lesssim\chi^2\lesssim17$ (which corresponds to a CL between $3$
and $4\sigma$ for 2 d.o.f.). Therefore, the selected realizations
are all in perfect agreement with current data. Note that one might naively expect to
find realizations with $\theta_{13}$ around the best-fit value of
$0^\circ$ since this value has been used as an input for
$M_\text{eff}^\text{exp}$. However, in almost all cases one obtains
$\theta_{23}$ to be around $50^\circ$ despite of the best-fit input value of $45^\circ$.

\begin{figure}[t]
\begin{center}
\includegraphics[width=8cm]{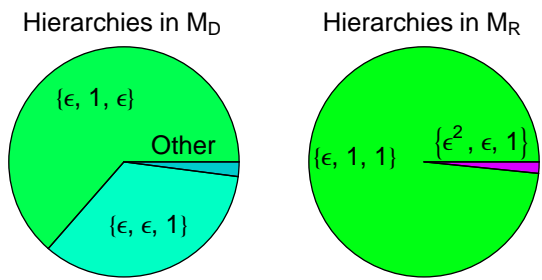}
\end{center}
\caption{\label{fig:mnorm} \footnotesize Distributions of hierarchies in $M_D$ (left) and $M_R$ (right)
leading to a NH neutrino masses for all valid seesaw realizations.}
\end{figure}

\figu{mnorm}, shows the distribution of mass spectra or hierarchies proportional to
$(m_1^D,m_2^D,m_3^D)$ and $(m_1^R,m_2^R,m_3^R)$ for $M_D$ and
$M_R$ respectively (each normalized to the corresponding heaviest mass eigenvalue
of $M_D$ and $M_R$) for the NH case. These distributions are obtained
by simply counting the number of realizations with a certain mass
spectrum. Observe that $M_R$ has as a mass spectrum only
$(\epsilon^2,\epsilon,1)$ and $(\epsilon,1,1)$. Note that the spectra labeled as
$(\epsilon^2,\epsilon,1)$ include also strongly hierarchical cases
where the right-handed Majorana neutrino mass spectrum can
actually be $(\epsilon^n,\epsilon,1)$, with some suitable $n\geq 3$
(see Appendix \ref{sec:paramhier}). \figu{mnorm} shows that, for
$M_R$, in all valid seesaw realizations, the mildly hierarchical mass
spectrum $(\epsilon,1,1)$ clearly
dominates the (strongly) the hierarchical spectrum
$(\epsilon^n,\epsilon,1)$ with $n\geq 2$. As it will turn out later
in Sec.~\ref{sec:nhtextures}, this observation is also supported at the texture-level: more than 80\% of
the extracted textures lead to a mild mass hierarchy $(\epsilon,1,1)$
for the right-handed neutrino masses. We
will come back to this point in the next paragraph. The absence of a degenerate mass spectrum for $M_R$ in the NH (and IH) case is a simple
consequence and selection effect of \equ{powers} in combination with
the assumption $M_\text{eff}^\text{exp}\sim\text{diag}(\epsilon^2,\epsilon,1)$. There are many more possibilities for the mass spectra of $M_D$ but they
are dominated by the types $(\epsilon,1,\epsilon)$ and
$(\epsilon,\epsilon,1)$. In \figu{mnorm}, the pie piece ``Other''
also contains the hierarchy $(\epsilon^2,\epsilon,1)$, which implies
that we have the same hierarchy in $M_R$ (but not vice versa). In our
method, no charged-lepton or quark-type hierarchy is produced.
We find from \figu{mnorm} that there are many possibilities to obtain a
normal neutrino mass hierarchy, but one cannot claim that this
hierarchy appears {\it typically} in $M_D$ or $M_R$ and then translates into $M_{\text{eff}}$.

The distributions of mass spectra in \figu{mnorm} may have immediate
relevance for leptogenesis \cite{leptogenesis} when crudely
extrapolating our results to the CP non-conserving case. In at least
80\% of the cases that we found (\cf, \Sec~\ref{sec:nhtextures}), the right-handed neutrino mass spectrum is of the mildly
hierarchical form $(\epsilon,1,1)$.  Thus, if the mass $m_1^R$ of the
lightest right-handed neutrino is in the range $m_1^R\lesssim
10^{12}\:\text{GeV}$, the seesaw scale set by
the mass of the heaviest right-handed neutrino $m_3^R$ would have
to be significantly lower than the usual $B-L$ breaking scale $\sim
10^{14}\:\text{GeV}$. For the mildly hierarchical right-handed
neutrino mass spectrum $(\epsilon,1,1)$, successful leptogenesis
might be achieved in two ways: (i) via resonant leptogenesis
\cite{resleptogenesis} (for recent models see, {\it e.g.},
Ref.~\cite{resonantmodels}) or (ii) by taking flavor effects into
account \cite{flavorleptogenesis} (for a connection with low-energy CP-violation
see, {\it e.g.}, Ref.~\cite{Pascoli:2006ie}). In the resonant limit, $m_3^R$ could be as low
as several TeV, thereby making this scenario testable at a
collider. Strongly hierarchical right-handed neutrino masses, which is
the standard case considered in the literature for leptogenesis,
are in our analysis found to be by about a factor of 5 less abundant
than the mild hierarchy. The possible strongly hierarchical
right-handed neutrino mass spectra are all of the type $(\epsilon^n,\epsilon,1)$, where $n\geq 2$. Allowing $n$ to be sufficiently large (say
$n=8$), the strongly hierarchical case can fit into a scheme with a seesaw scale of the order $m_3^R\sim
10^{14}\:\text{GeV}$ and sufficient baryon asymmetry could again be
generated through flavored leptogenesis.

\begin{figure}[t]
\begin{center}
\includegraphics[width=\textwidth]{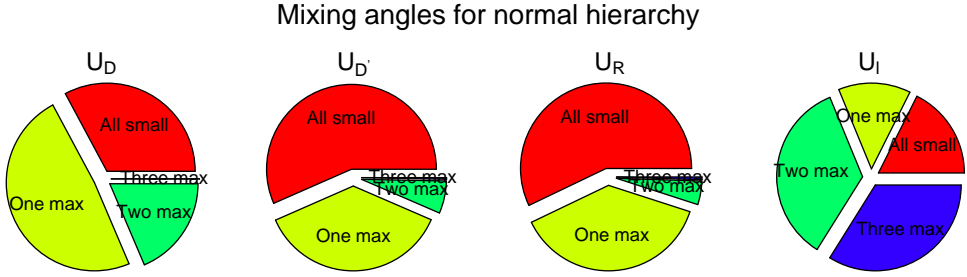}
\end{center}
\caption{\label{fig:mixnormsum}  \footnotesize Distributions of mixings in $U_D$, $U_{D'}$, $U_R$, and $U_\ell$ (in columns) leading to a normal neutrino mass hierarchy. The different pie labels refer to the number of maximal mixing angles, where ``All small'' corresponds to all mixing angles $\le\epsilon$.}
\end{figure}

\figu{mixnormsum} shows the distributions of mixing angles, where we concentrate on the number of maximal
(``max'') mixing angles $\theta_{ij}^x=\pi/4$ appearing in $U_D$,
$U_{D'}$, $U_R$, and $U_\ell$. If there is no maximal mixing angle, we call the scenario ``All small``. Let us first note that
the distributions of the mixing angles in $M_\nu$ are very different
from $U_\ell$. In $U_\ell$, we often find large mixings, which means
that the large lepton mixing angles are not necessarily created in the
neutrino sector, but can also come very often from the charged lepton
sector. Note that the pie slice ``All small'' in $U_\ell$ represents
more or less CKM-like mixings in $U_\ell$.
There are also many possibilities with ``trimaximal''
mixing\footnote{Not to be confused with tri-bimaximal mixing.} ({\it i.e.},
all three mixing angles $\theta_{ij}^x$ are maximal for some given
sector $x$) in $U_\ell$. This is different from the other mixing
matrices, where trimaximal mixing hardly occurs, and either one
maximal mixing angle or only small mixing angles are preferred. In
particular, in $U_{D'}$ and $U_R$, only small mixings are typical.

\begin{figure}[t]
\begin{center}
\includegraphics[width=13.5cm]{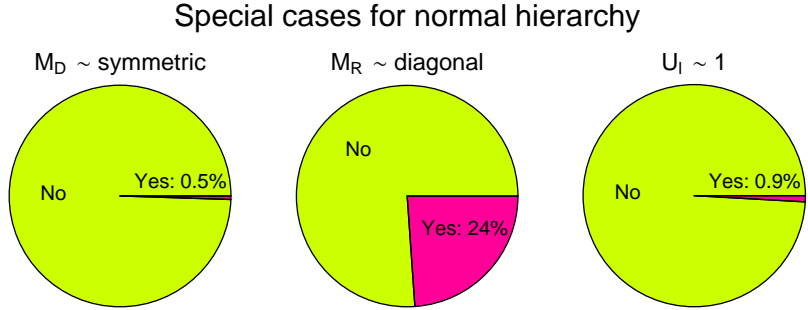}
\end{center}
\caption{\label{fig:special} \footnotesize Fraction of special cases for normal mass hierarchy. 
The different pies show the fractions of the cases with symmetric
$M_D$, diagonal $M_R$, and $U_\ell \simeq \mathbbm{1}$, of all allowed realizations (not texture sets).
This figure is based on the mixing matrices, where $U_D \simeq U_{D'}$ in the first case,
$U_R \simeq \mathbbm{1}$ and $U_\ell \simeq \mathbbm{1}$ in the second and third case, respectively. For the
similarity condition ``$\simeq$'', we allow $\epsilon^2$-deviations in the mixing angles.
For instance, for an exact $U_R=\mathbbm{1}$, one would have only 2\% of all realizations.}
\end{figure}
\figu{special} shows the fraction of realizations that exhibit symmetric
 $M_D$, and/or diagonal $M_R$, and/or $U_\ell \simeq \mathbbm{1}$.
In this figure, ``$M_D\sim$ symmetric'' means that $M_D$ is symmetric
 up to possible corrections of the order $\epsilon^2$. It is evident
 that there are only very few realizations with symmetric $M_D$ or $U_\ell \simeq
\mathbbm{1}$, and none with diagonal $M_D$,  which is not surprising from
what we have learned above. One may conclude from this result that
there are plenty of possibilities to implement the seesaw mechanism
 without these constraints, which have, however, often been imposed in
 existing literature.

Generally, note that, while one may argue that one can construct more possibilities in a more general and
sophisticated framework, our realizations result from very generic
and simple assumptions without adding another level of complexity. In
this sense, our generic assumptions are much simpler than ad-hoc constraints,
such as requiring that $U_\ell$ be diagonal. It is interesting to note that we find for $M_R$ quite many realizations that are close to
a diagonal form.

\section{A Selection of Textures for the NH Case}
\label{sec:nhtextures}

In this section, we present a selection of texture sets
for the NH case. These texture sets satisfy certain selection criteria listed in \App~\ref{sec:texselection}. 
The full set of the texture sets is available in \Ref~\cite{TexWeb}.
\Tab~\ref{tab:seesawtextures}  shows these 72 texture sets for $M_\ell,M_D,$ and $M_R$, together with associated example realizations leading to these texture sets. For each texture set, we always choose the realization with the lowest $\chi^2$, \ie, the realization which fits data best. For this realization, the table lists the mass  spectra of $M_D$ and $M_R$, as well as the mixing angles $\theta^x_{ij}$  in the different sectors. For each case in \Tab~\ref{tab:seesawtextures}, we have collected in \Tab~\ref{tab:seesawtexturesphases} (in \App~\ref{sec:realizationsrest}) all the corresponding phases ($0$ or $\pi$) in order to allowing for a complete reconstruction of the realizations and Yukawa coupling matrices. In addition, one can find there the PMNS mixing angles, as well as the number of realizations that become
 identified with each texture set through the texture reduction. In \Tab~\ref{tab:seesawtextures}, the parameter $\xi$ can take the values
 $\xi\in\{0,\epsilon^2\}$.


\begin{footnotesize}
\begin{center}
\begin{longtable}{|@{\hspace{1mm}}c@{\hspace{1mm}}||@{\hspace{0.5mm}}c@{\hspace{0.5mm}}|@{\hspace{0.5mm}}c@{\hspace{0.5mm}}|@{\hspace{0.5mm}}c@{\hspace{0.5mm}}||@{\hspace{0.5mm}}c@{\hspace{0.5mm}}|@{\hspace{0.5mm}}c@{\hspace{0.5mm}}|@{\hspace{0.3mm}}c@{\hspace{0.3mm}}|@{\hspace{0.3mm}}c@{\hspace{0.3mm}}|}
\hline &&&&& \vspace{-3mm} \\*
\# & $M_\ell$ & $M_D$ & $M_R$ & 
$\begin{array}{@{\hspace{0.5mm}}c@{\hspace{0.5mm}}}
 m_i^D/m_D \\[1mm]
 m_i^R/M_{B-L}
\end{array}$
 & 
$\begin{array}{@{\hspace{0.5mm}}c@{\hspace{0.5mm}}}
 (\theta _{12}^\ell,  \theta _{13}^\ell, \theta _{23}^\ell) \\[1mm]
 (\theta _{12}^D, \theta _{13}^D, \theta _{23}^D) \\[1mm]
 (\theta _{12}^{D'}, \theta _{13}^{D'},  \theta _{23}^{D'}) \\[1mm]
 (\theta _{12}^R, \theta _{13}^R,  \theta _{23}^R) 
\end{array}$
\\*[1mm] \hline

 $1$ & $
\left(
\begin{array}{ccc}
 0 & 0 & \epsilon  \\
 0 & \epsilon ^2 & 1 \\
 0 & \epsilon ^2 & 1
\end{array}
\right)$ & $\left(
\begin{array}{ccc}
 \epsilon ^2 & \epsilon  & 0 \\
 \xi  & 1 & 1 \\
 \epsilon ^2 & \xi  & \epsilon 
\end{array}
\right)$ & $\left(
\begin{array}{ccc}
 \epsilon ^2 & 0 & 0 \\
 0 & 1 & 1 \\
 0 & 1 & 1
\end{array}
\right)$ & $
\begin{array}{c}
 ( \epsilon ^2 , 1 , \epsilon  ) \vspace{0.1cm}\\
 ( \epsilon ^2 , \epsilon  , 1 )
\end{array}
$ & $
\begin{array}{c}
 ( \xi  , \epsilon  , \frac{\pi }{4} ) \\
 ( \epsilon  , \frac{\pi }{4} , \xi  ) \\
 ( \xi  , \xi  , \frac{\pi }{4} ) \\
 ( \xi  , 0 , \frac{\pi }{4} )
\end{array}
$

\\
\end{longtable}
\end{center}
\end{footnotesize}
In \Tab~\ref{tab:classification}, we divide some of the interesting
 textures/realizations from \Tab~\ref{tab:seesawtextures} into certain
 classes, such as lopsided $M_\ell$ (only $\theta_{23}^\ell$ is maximal),
 anarchic\footnote{Cases of anarchic $M_R$ cannot appear in
   \Tab~\ref{tab:seesawtextures} due to the selection criteria in
 \App~\ref{sec:texselection}.} $M_D$
 ($\theta_{13}^{D}=\theta_{23}^{D}=\theta_{13}^{D'}=\theta_{23}^{D'}=\frac{\pi}{4}$),
 lopsided $M_D$ (at least one of the mixing angles $\theta_{12}^{x}$
 and $\theta_{23}^{x}$, $x=D,D'$, is maximal -- anarchic cases
 are excluded), hierarchical $M_R$ (hierarchical mass
 spectrum but no maximal mixing angle for $M_R$, \ie,
 $M_R^\text{diag}\propto\text{diag}(\epsilon^2,\epsilon,1)$),
 semi-anarchic $M_R$ (one of the angles $\theta_{12}^R$ or
 $\theta_{23}^R$ is maximal), ``diamond''-type $M_R$ ($\theta_{13}^R=\pi/4$
 and the corresponding texture has a diamond shape~\cite{Plentinger:2006nb}), and presence of a ``dead
 angle $\xi$'' (a mixing angle $\xi\lesssim\epsilon^2$ that does not
 affect the corresponding matrix in the texture set).\footnote{For example,
 if a matrix element reads
 $A\,\epsilon^2+B\,\xi\epsilon+\mathcal{O}(\epsilon^4)$, where $A$ and
 $B$ are order one coefficients, then the choice of
 $\xi\in\{0,\epsilon^2\}$ has no impact on the texture, \ie,
 $A\,\epsilon^2+\mathcal{O}(\epsilon^4)\rightarrow\epsilon^2$ and
 $A\,\epsilon^2+B\,\epsilon^3+\mathcal{O}(\epsilon^4)\rightarrow\epsilon^2$.} Note
 that, as already mentioned in Sec.~\ref{sec:nhfull}, about 80\% of all
 textures in \Tab~\ref{tab:seesawtextures} have a mildly hierarchical
 spectrum for the right-handed neutrino masses that is of the form
 $M_R^\text{diag}\propto\text{diag}(\epsilon,1,1)$, whereas a strongly
 hierarchical spectrum
 $M_R^\text{diag}\propto\text{diag}(\epsilon^n,\epsilon,1)$ ($n\geq
 2$) occurs in only roughly 20\% of the cases. Although
 this classification may to a certain extent be incomplete, it could 
nevertheless serve to characterize significant features of the
 textures. Moreover, all realizations in \Tab~\ref{tab:seesawtextures} have in common that
$\theta_{12}\simeq 33^\circ$, $\theta_{13}\simeq 0^\circ$, and
$\theta_{23}\simeq 51^\circ$ (see
\Tab~\ref{tab:seesawtexturesphases}). Therefore, if future experiments
measure $\theta_{23}$ smaller than $45^\circ$, the presented
 realizations could be tested.

\begin{table}[t]
\begin{center}
\begin{tabular}{ll}
\hline
Class & Texture \# \\\hline \hline
$U_\ell \simeq V_{\text{CKM}}$ & 46, 47\\
$U_\ell\simeq \mathbbm 1$ & 33, 34, 71, 72\\
$\theta_{ij}^\ell\neq\frac{\pi}{4}$ & 20, 27, 33--35, 44, 46, 47, 49--53, 57, 58, 62--64,\\
 & 67, 68, 71, 72\\
Lopsided $M_\ell$  & 
1, 4--8, 10, 11, 13, 43, 48, 56, 60, 61, 69, 70 \\
Bimaximal$M_\ell$  & 9, 17, 21--23, 26, 32, 38--40, 55, 59, 65,  \\
Trimaximal $M_\ell$  & 3, 12, 14, 15, 16, 18, 19, 25, 28 \\ \hline
$U_D\simeq U_{D'}$ & 12, 23\\
Anarchic $M_D$ & 24, 54 \\ 
Lopsided $M_D$ & 8, 20, 21, 23, 27, 32, 33, 35, 41, 42, 45, 49,\\
 &  50, 52, 53, 62--64, 67, 68 \\ \hline 
$U_R\simeq \mathbbm 1$ & 10, 33\\
Hierarchical $M_R$ & 48, 58, 65, 66 \\
Semi-anarchic $M_R$ & 1, 16, 18, 24, 32, 46, 47, 49, 50, 54--57, 60--63 \\ 
Diamond $M_R$ & 2, 3, 5, 9, 15, 17, 19, 29-31, 35, 43, 44, 67-72 \\ \hline 
$M_D^\text{diag}\propto M_R^\text{diag}~$ & 43, 44, 48, 58, 65, 66, 71, 72 \\
$M_R^\text{diag}\propto\text{diag}(\epsilon^2,\epsilon,1)~$
& 1, 43, 44, 48, 49, 50, 58, 61, 62, 63, 65, 66, 71, 72 \\\hline
Dead angle $\xi$ &  1, 6, 7, 9, 10, 13, 14, 24, 28, 30--33, 36--38, 41, 42,\\
 & 46--48, 52, 53, 57, 61--64, 68, 70--72 \\
\hline
\end{tabular}
\end{center}
\caption{\footnotesize Classification of textures and realizations
  which may be of special interest. Here, ``$\simeq$'' means up to
  phases. The case $M_D^\text{diag}\propto M_R^\text{diag}$ allows,
  in combination with \equ{powers}, only for the mass spectra
  $(\epsilon^2,\epsilon,1)$. Note that the relative fraction of hierarchical
  $M_R^\text{diag}$ in \Tab~\ref{tab:seesawtextures} is much higher
  than prior to applying the selection criteria, as it is evident from \figu{mnorm}.
}
\label{tab:classification}
\end{table}

Let us now illustrate how the textures in
\Tab~\ref{tab:seesawtextures} could be generated in explicit models by
considering two examples. For
this purpose, assume an $M$-fold $Z_N$ product flavor symmetry group
$G_F=\Pi_{i=1}^M Z_N^{(i)}=Z_N\times Z_N\times\dots\times Z_N$. We suppose that for each
individual group $Z_N^{(i)}$ there are two types of SM singlet flavon
fields $f_i$ and $f_i'$ that carry different $Z_N^{(i)}$ charges but which
are singlets under transformations of all the other groups $Z_N^{(j)}$, with $j\neq
i$. All flavons shall acquire universal VEVs: $\langle f_i\rangle\simeq\langle f_i'\rangle\simeq v$, for $i=1,2,\dots, M$, where
$v=\epsilon\cdot M_F$ and $M_F$ is some fundamental Froggatt-Nielsen
(see Sec.~\ref{sec:GUTs}) messenger scale. Now, let us specialize to the case $M=7$ and
$N=4$, and assume that each pair of flavons $f_i$ and $f_i'$ carries the
$Z_4^{(i)}$ charges $f_i\sim 1$ and $f_i'\sim 2$, {\it i.e.}, $f_i$ is
singly and $f_i'$ doubly charged under $Z_4^{(i)}$. Technically, this
amounts to realizing fractional charges for the $Z_2$-subgroups of
$Z_4^{(i)}$. We assign the leptons $G_F$ quantum numbers as shown in
\Tab~\ref{tab:ZNcharges} for two example models.
\begin{table}
\begin{center}
\begin{tabular}{c||c||c}
\hline
Field & Model 1& Model 2\\
\hline 
$\nu^c_1$ & (0,0,0,1,0,1,1) &  (2,0,0,2,0,0,1) \\ 
$\nu^c_2$ & (2,0,0,1,0,1,1) &  (2,0,0,2,0,0,1) \\ 
$\nu^c_3$ & (0,0,0,1,0,1,1) &  (0,2,0,0,2,1,0)\\
\hline 
$\ell_1$ & (0,2,0,0,1,0,1) &  (2,0,2,2,2,1,0) \\ 
$\ell_2$ & (0,0,0,0,1,1,0) &  (2,2,0,2,2,1,0)\\ 
$\ell_3$ & (0,0,2,1,0,0,1) &  (0,2,2,2,2,0,1) \\
\hline 
$e^c_1$ &  (2,2,2,1,1,1,1) & (0,0,0,0,0,1,1)\\ 
$e^c_2$ &  (2,2,2,0,0,1,0) & (2,2,2,0,0,3,3)\\ 
$e^c_3$ &  (0,2,2,0,0,0,0) & (2,2,2,2,2,3,3)\\ 
\hline
\end{tabular}
\end{center}
\caption{\label{tab:ZNcharges}\footnotesize Assignment of $G_F$ charges to the
  leptons. Models 1 and 2 lead respectively to the textures \#17 and
  18 in \Tab~\ref{tab:seesawtextures} (see text).}
\end{table}
In both models, we have assigned in \Tab~\ref{tab:ZNcharges} each lepton a row vector $(q_1,q_2,\dots,q_7)$, where $q_i$ denotes the $Z_4$
charge of the lepton under the group $Z_4^{(i)}$
($i=1,2,\dots,7$). Models 1 and 2 in
\Tab~\ref{tab:ZNcharges} respectively produce the texture sets
\#17 and 18 in \Tab~\ref{tab:seesawtextures} via the Froggatt-Nielsen
mechanism. Since we have already found in
\Tabs~\ref{tab:seesawtextures} and \ref{tab:seesawtexturesphases}
valid realizations for these textures, we know, without any further
calculation, that the order unity Yukawa couplings in model 1 and
model 2 can be chosen such that they reproduce the lepton mass and mixing parameters in perfect agreement
with data. Indeed, the explicit realizations in
\Tabs~\ref{tab:seesawtextures} and \ref{tab:seesawtexturesphases}
allow, if one wishes, for a complete reconstruction of such a valid
set of order one Yukawa couplings.

The two models above represent only very specific examples of how one
could directly apply \Tab~\ref{tab:seesawtextures} to identify the
possible flavor symmetries and their breaking in a model. However, there are certainly many
more possibilities. For example, we assumed here, for simplicity,
copies of a single discrete Abelian flavor symmetry group, but new possibilities
arise when considering product groups with different $Z_N$ subgroups
or also non-Abelian flavor symmetries. Moreover, note that, since $U_{\ell'}$ has been
set equal to the unit matrix, \Tab~\ref{tab:seesawtextures} lists only a few percent of the actual
total number of texture sets in extended QLC. A systematized scan for models
generating the textures in extended QLC should also include these
extra cases. In addition, it would be interesting to explore the
compatibility with further constraints from GUT-relations or cancellation of anomalies (see, {\it e.g.}, \Ref~\cite{Batra:2005rh}).

\section{IH and QD Case}
\label{sec:invdeg}

Even though the main focus of our study is the case of NH neutrino
masses, we have also calculated the valid realizations for IH and QD
neutrinos using \equ{numassratios} for the respective mass
hierarchies. These cases are qualitatively different from the NH case
because RG effects may be relevant here.
In this section, we present a qualitative discussion of the IH and QD cases.

\subsection{IH Spectrum}
\label{sec:inverted}

In the IH case, we find $797 \, 928$ different realizations,
which corresponds to $10 \, 196$ different qualitative cases ignoring
phases. This is about a factor of five more realizations than in the
NH case. After texture reduction, one obtains $4 \, 268$ different texture sets,
\ie, different combinations of $M_\ell$, $M_D$, and $M_R$, which is about twice the number as in
the NH case. This indicates a higher redundancy at the level
of the realizations as compared to the NH case,
which we will comment on more explicitely in the QD case in \Sec~\ref{sec:degenerate}.
Since some of the selection criteria for NH neutrino masses are based
on the $\chi^2$ selector, which is not defined in the IH case, the
texture reduction of the texture sets for the IH case has to be carried
out in a way that differs from the procedure described in Appendix \ref{sec:texselection}.
However, by arguments similar to those in Appendix
\ref{sec:texselection}, we find a reduction of the number of cases
that is comparable to that in the NH case.

\begin{figure}[t]
\begin{center}
\includegraphics[width=8cm]{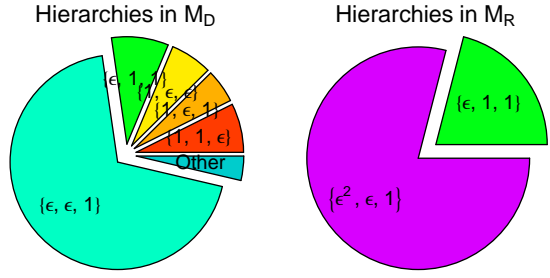}
\end{center}
\caption{\label{fig:minv} \footnotesize Distributions of mass
  hierarchies for $M_D$ (left) and $M_R$ (right)
leading to IH neutrino masses for all seesaw realizations.}
\end{figure}

The mass spectra of $M_D$ and $M_R$ for the IH case are shown in
\figu{minv}. In contrast to the case of NH neutrino masses
(\cf, \figu{mnorm}), $M_D$ shows a larger variation of
possible mass hierarchies. It turns out that most realizations for $M_D$ have a
mass hierarchy of the type $(\epsilon,\epsilon,1)$. In $M_R$, we
obtain the same mass spectra as for the NH case, but with a nearly
opposite weighing. For $M_D$, one may have an inverse mass hierarchy,
whereas $M_R^\text{diag}$ has by construction a ``normal'' ordering
(\cf, Appendix \ref{sec:paramhier}).

\begin{figure}[t]
\begin{center}
\includegraphics[width=\textwidth]{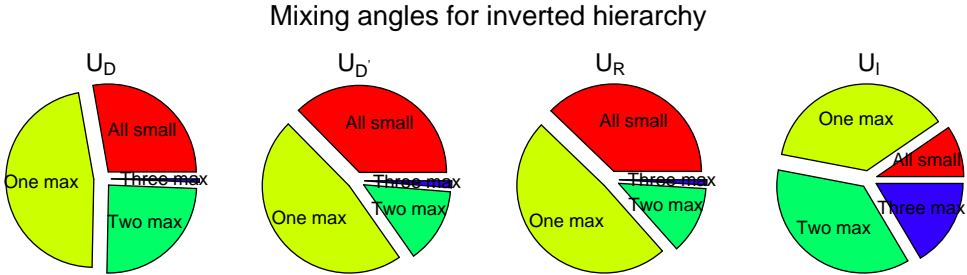}
\end{center}
\caption{\label{fig:mixinvsum} 
\footnotesize Distributions of mixing angles in $U_D$, $U_D'$, $U_R$,
and $U_\ell$ (in columns) that lead to IH neutrino masses for all seesaw realizations.  The different pie labels refer to the number of maximal 
mixing angles, where ``All small'' corresponds to all mixing angles $\leq\epsilon$.}
\end{figure}

\figu{mixinvsum} shows the distribution of maximal mixing angles for
the IH case. Note that one observes a similar distribution for NH
neutrinos (\cf~\figu{mixnormsum}), namely that in most cases
large leptonic mixing angles come from $U_\ell$. However, for IH neutrino masses,
there is a slight tendency to larger mixing angles in $U_D$, $U_{D'}$,
and $U_R$, and smaller mixing angles in $U_\ell$ than in the NH
case. The distribution of the special cases in \figu{special} resembles
that for the normal hierarchy. For example, $M_R$ is approximately diagonal in about 15\% of all cases.

\subsection{QD Spectrum}
\label{sec:degenerate}

The case of QD neutrino masses is more tricky from the computational point of view
because of redundancy. As a simple example, consider \equ{meffexp} with $M_{\text{eff}}^{\text{diag}} = \mathbbm{1}$. In the real case, one can use any $\widehat{U}_{\text{PMNS}}$ and $\widehat{U}_{\ell}$ in this equation
in order to obtain $M_{\text{eff}}^{\text{exp}} = \mathbbm{1}$, \ie,
the different cases for $\widehat{U}_{\ell}$ are not really
qualitatively different and a statistical analysis, such as above,
does not make much sense. In addition, the matching to experiment (or the experimental data as a selection criterion) is, in this way, rather meaningless. From \equ{matching}, one can read off that we require $M_{\text{eff}}^{\text{th}} \propto \mathbbm{1} + \mathcal{O}(\epsilon^3)$.  This implies that 
any realization producing $M_{\text{eff}}^{\text{th}} \simeq c \cdot
\mathbbm{1}$ will be accepted by the algorithm, which are quite many.
An interesting observation in the degenerate case, is that the
realizations divide into two qualitatively different cases: the trivial case, \ie, $M_D = M_R = \mathbbm{1}$, and a large
number of non-trivial cases. There is no such trivial case for IH
neutrinos.

\section{Summary and Conclusions}
\label{sec:summary}

Quark-lepton unification generally leads to predictions
in the patterns of the Yukawa coupling matrices of the fermions, which are often called ``textures''.
Because of the high dimension of the parameter space,
a direct systematic study or scan of all valid textures at the level of the
Yukawa couplings is difficult to perform. Therefore,
we suggest a ``bottom up'' search strategy: We construct the parameter space of
all valid textures from very generic (low-level) assumptions in the
relevant leptonic mixing matrices
and mass spectra. In this approach, we use the context of extended quark-lepton complementarity, which means that all mixing angles are either maximal or 
described by powers of $\epsilon \simeq \theta_\text{C}$. In addition,
all mass ratios (of the neutrinos and charged leptons, as well as the
mass hierarchies in $M_D$ and $M_R$) are also parameterized by powers
$\epsilon$. This expansion parameter may be the remnant of a
flavor symmetry describing the masses and mixing angles in both the
quark and lepton sectors within a quark-lepton unified context. Note
that in this setting, the solar neutrino mixing angle can only emerge
as a combination of maximal mixing and $\theta_\text{C}$, \ie, our assumptions are somewhat more
general than the usual QLC relations [see, {\it e.g.}, \equ{qlc}].

Our procedure, in short terms, is as follows (\cf, \figu{structure}): First, we 
systematically construct all combinatorial possibilities for all
mixing matrices and mass spectra for the type-I seesaw mechanism up to
the order $\epsilon^2$. In each case, we compute the effective neutrino
mass matrix $M_{\mathrm{eff}}^{\text{th}}$ from that.
In the second step, we determine all possibilities for $M_{\mathrm{eff}}^{\text{exp}}$
containing the experimental information from $U_{\text{PMNS}}$, the neutrino mass hierarchy, and $U_\ell$.
In the third step, the two matrices are matched, \ie, the realizations compatible
with current data are selected. Finally, we identify the leading order entries in the matrices,
which lead to the sets of textures. We scan about 20 trillion possibilities focusing on the
real, CP conserving case, and we mainly discuss a normal neutrino mass
hierarchy. As an experimentally motivated input value for
$\stheta$, we take the current best-fit value $\stheta=0$ as an example.

Compared to individual models in the literature, our assumptions are too generic 
to be able to predict the outcome {\it a priori}, which may reduce the bias. 
 This key feature allows for the  interpretation of the valid 
result statistics with respect to observables, constructed hierarchies, \etc. 
For example, we find many different cases with a very mild hierarchy
in $M_R$ (with an abundance of about 80\% in the textures),
a result which does not involve any bias from the
point of view of the assumptions. In addition, we find only very few
cases with a symmetric Dirac neutrino mass matrix $M_D$ or small mixing
($U_\ell \simeq \mathbbm{1}$) in the charged lepton sector.
However, we find a roughly diagonal right-handed neutrino mass matrix
$M_R$ in relatively many
cases, \ie, in about 24\% of all realizations for the normal neutrino mass hierarchy.
It would be interesting to investigate 
signals of lepton flavor violation for our list of matrices, and connect our results to
statistical studies along the lines of Ref.~\cite{anarchy} or to
recent attempts of scanning the SM parameters including quark and lepton masses \cite{statmasses}.

We have presented a complete selection of 72 examples for Yukawa coupling
 textures satisfying specific selection
criteria. A more complete list of 1981 texture sets (obtained by
relaxing the selection criteria) can be found in
 \Ref~\cite{TexWeb}. We have shown in two examples that the textures
 are very useful for a direct search for flavor symmetries predicting
 current data. 
Therefore, our list(s) of textures can be understood
as an intermediate result from the point of view of model building.
They might be used to systematically test and find flavor symmetries
and allow for a more systematic approach to constructing models.

We conclude that systematic, machine-supported searches of parameter spaces related
to model building may, in fact, open up new possibilities. While conventional approaches
focus on individual models in greater depth, our approach produces
{\em all} valid lepton mass matrices with the only bias of the (often very generic) input assumptions.
Therefore, they allow for the identification of new possibilities, as well as 
for more general studies of the discussed parameter space. Hence, our study should not only
be interpreted with respect to the input assumptions, but also with respect to the
procedure itself.

\section*{Acknowledgments}

We would like to thank Wilfried Buchm\"uller, Laura Covi, Alexandro
Ibarra, J{\"o}rn Kersten, Hitoshi Murayama, Tommy Ohlsson, Serguey Petcov, Werner Porod, Thomas Schwetz,
and Alexei Smirnov for useful discussions. The research of F.P. is supported by Research Training Group 1147 \textit{Theoretical Astrophysics and Particle Physics} of Deutsche Forschungsgemeinschaft.
G.S. was supported by the Federal Ministry of Education and
Research (BMBF) under contract number 05HT1WWA2.
W.W. would like to acknowledge support from the Emmy Noether program
of Deutsche Forschungsgemeinschaft.

\appendix

\section{Details of the Procedure, Heuristics, and Complexity}
\label{sec:relevant}

In this appendix, we give details of the algorithm, heuristics (also physically relevant ones, everything
which makes the algorithm faster but does not lead to a loss of generality), and complexity.

\subsection{Parameter Space for the Mass Hierarchies}
\label{sec:paramhier}

In the procedure described in \Sec~\ref{sec:generating}, we have considered all
possible diagonal mass matrices
$M_D^\text{diag}=m_D\,\text{diag}(\epsilon^a,\epsilon^b,\epsilon^c)$ and
$M_R^\text{diag}=m_D\,\text{diag}(\epsilon^{a'},\epsilon^{b'},\epsilon^{c'})$,
with suitable non-negative integers $a,b,c,a',b',$ and $c'$, that are
compatible with the observed neutrino mass squared differences and
leptonic mixing angles. As we will now see, the relevant range of these integers
can actually be restricted considerably if we take the
parameterization of the neutrino mass spectrum in \equ{numassratios} into account.

By factoring out common powers in $\epsilon$, we see
that $M_D^\text{diag}$ and $M_R^\text{diag}$ can be written as
\begin{subequations}\label{equ:mdmrpowers}
\begin{equation}
M_D^{\mathrm{diag}}  = 
m_D\,\text{diag}(\epsilon^m,\epsilon^n,1),\quad
M_D^{\mathrm{diag}}  = 
m_D\,\text{diag}(\epsilon^m,1,\epsilon^n),\quad
M_D^{\mathrm{diag}}  = 
m_D\,\text{diag}(1,\epsilon^m,\epsilon^n),
\label{equ:mddiag}
\end{equation}
and
\begin{equation}
M_R^\text{diag}  =  M_{B-L} \,
\mathrm{diag}(\epsilon^p,\epsilon^q,1)\:\:
\text{with}\:\:0\leq q\leq p, 
\label{equ:mrdiag}
\end{equation}
where $m,n,p$, and $q$, are non-negative integers, and
$m_D\sim 10^2\:\text{GeV}$. In \equ{mrdiag}, we have made use of the
possibility to bring $M_R^\text{diag}$ to a strictly hierarchical form, which is
expressed by the condition $0\leq q\leq p$. One then has, however, no
longer the freedom to choose the order of the mass eigenvalues of
$M_D^\text{diag}$, and that is why \equ{mddiag} includes all
permutations of $M_D^\text{diag}$ without any specific ordering of
$m$ and $n$, {\it i.e.}, we allow in \equ{mddiag} for both the cases $n\leq m$ and
$n>m$. In the following, we set $m_\nu\equiv
m_D^2/M_{B-L}$. Although $m_D\sim
10^{2}\:\text{GeV}$ and $M_{B-L}\sim 10^{14}\:\text{GeV}$ would be
preferred values, we can, of course, always
rescale $m_D\rightarrow a\cdot m_D$ and $M_{B-L}\rightarrow
a^2\cdot M_R$ by some factor $a$, thereby leaving the absolute neutrino mass
scale $m_\nu\sim 10^{-2}\:\text{eV}$ unchanged. From
$\text{det}\,M_\text{eff}^\text{diag}=(\text{det}\,M_D^\text{diag})^2\text{det}\,(M_R^\text{diag})^{-1}$
we find the relation
\begin{equation}\label{equ:powers}
 2(m+n)-p-q=k,
\end{equation}
\end{subequations}
where $k$ is a non-negative integer that depends on the type of
neutrino mass spectrum in \equ{numassratios}: We have $k=3$ for a normal, $k=1$
for an inverted, and $k=0$ for a degenerate neutrino mass
spectrum.\footnote{From \equ{powers} and $p\geq q$, it also follows
  that $m,n\leq\frac{k}{2}+p$.}

We are now interested in determining the allowed ranges for the integers
$m,n,p,$ and $q$, in Eqs.~(\ref{equ:mdmrpowers}). Later, we will apply
\equ{matching} up to order $\epsilon^2$, \ie, we will require a numerical
matching precision $\mathcal{O}{(\epsilon^3)}$ between $M_{\text{eff}}^{\text{th}}$ and 
$M_{\text{eff}}^{\text{exp}}$. From that matching precision it follows that
only powers up to $\epsilon^2$ are relevant in the individual factors in the product \equ{meffth},
because higher order terms will be absorbed
by this matching uncertainty. The only non-trivial aspect in this argument is the
factor $M_R^{-1}$. 
Let us write $(M_R^\text{diag})^{-1}=\epsilon^{-p}M_{B-L}^{-1}\,
\text{diag}(1,\epsilon^{p-q},\epsilon^p)$, and redefine
$m_D\rightarrow\epsilon^{p/2}m_D$ and $M_R\rightarrow\epsilon^{p}M_R$, which leaves $m_\nu$
unaffected. Consider now the entries $\epsilon^m$, $\epsilon^n$, $\epsilon^{p-q}$,
and $\epsilon^p$, in $M_D^\text{diag}$ and $(M_R^\text{diag})^{-1}$. If
$m$, $n$, $p-q$, or $q$, are larger than 2, then the corresponding
contribution to $M_\text{eff}$ will be absorbed in the matching precision. 
It follows that it is 
sufficient to restrict the maximum values of $m,m,p,$ and $q$ to 2,
\ie, we need to consider for the powers of $\epsilon$ only
\begin{equation}
0\leq m,n,p,q\leq 2,
\label{equ:powerrange}
\end{equation}
where we have used that $0\leq q\leq p$. The matrices $M_D^\text{diag}$ and $M_R^\text{diag}$ with higher powers of $\epsilon$ fall then into one of the classes already covered by a combination of powers satisfying
\equ{powers}.

\subsection{Matching of Matrices and the Absolute Neutrino Mass Scale}
\label{sec:comparison}

In this section, we describe the numerical implementation of \equ{matching}:
 $M_\text{eff}^\text{th}|_{\epsilon=0.2} \simeq M_\text{eff}^\text{exp}|_{\epsilon=0.2}$.
For our procedure, it makes sense to require a matching precision of
 $\mathcal{O}(\epsilon^3)$  since $\epsilon^2$ is the maximum power
 that is used for the mass hierarchies and mixing angles (which was chosen
because higher orders would be absorbed by the current measurement precision).
This implies that a higher order matching precision 
will be too precise/restrictive because $\epsilon^3$ powers are not directly produced, and
a lower order matching precision would not be able to distinguish cases differing by $\epsilon^2$-terms.
Numerically, the $\mathcal{O}(\epsilon^3)$ term can have any order one
 coefficient. We require a precision of $P = \epsilon^3$ for \equ{matching}, which
turns out to be reasonably restrictive by using actual tests, \ie, we require\footnote{The dataset of 
valid models is very sensitive to this $P$. If it is much smaller, we
 do not find any valid realizations because $\epsilon^3$-terms are not
 directly produced by our procedure. 
If it is somewhat larger, the found realizations fail consistency with data. For example, for
a normal neutrino mass hierarchy, one can diagonalize $M_\text{eff}^\text{th}$ and read off the
mixing angles as in \Ref~\cite{Plentinger:2006nb}. These mixing angles can
be compared with the actual ones which have been used as input
 values for $M_\text{eff}^\text{exp}$. We
find that our matching precision
is sufficient such that only realizations compatible with current
data at least at the $5\sigma$ confidence level are allowed for the NH
 case (most of them actually
fit much better); \cf, \figu{allchi2}. 
A weaker constraint on $P$ allows more realizations which provide, however, a too bad fit.}
\begin{subequations}
\begin{equation}
(M_\text{eff}^\text{exp})_{ij} - P \le c \cdot (M_\text{eff}^\text{th})_{ij} \le (M_\text{eff}^\text{exp})_{ij} + P,
\label{equ:numprec}
\end{equation}
with $P = \epsilon^3$ and $c$ is a constant which can be absorbed into
(or can come from) the absolute mass scale, mixing angles, and the highest hierarchy power in $M_R$.

In order to evaluate  \equ{numprec}, we have to determine $c$ in order
to check whether $M_\text{eff}^\text{th}$ falls into the proper interval.
Let us re-write \equ{numprec} as
\begin{equation}
(M_\text{eff}^{\mathrm{exp}})_{ij}^\mathrm{low} \le c\cdot(M_\text{eff}^\text{th})_{ij} \le (M_\text{eff}^{\mathrm{exp}})_{ij}^\mathrm{up} \, .
\label{equ:vrange}
\end{equation}
 This corresponds to 6$\times$2 independent inequalities. Dividing now these inequalities by $(M_\text{eff}^\text{exp})_{ij}$ gives
\begin{eqnarray}
\frac{(M_\text{eff}^{\mathrm{exp}})_{ij}^\mathrm{low}}{(M_\text{eff}^\text{th})_{ij}} \le  & c & \le \frac{(M_\text{eff}^{\mathrm{exp}})_{ij}^\mathrm{up}}{(M_\text{eff}^\text{th})_{ij}} \quad \mathrm{for}
\, \, (M_\text{eff}^\text{th})_{ij} > 0 \, , \nonumber \\
\frac{(M_\text{eff}^{\mathrm{exp}})_{ij}^\mathrm{up}}{(M_\text{eff}^\text{th})_{ij}} \le  & c & \le \frac{(M_\text{eff}^{\mathrm{exp}})_{ij}^\mathrm{low}}{(M_\text{eff}^\text{th})_{ij}} \quad \mathrm{for}
\, \, (M_\text{eff}^\text{th})_{ij} < 0\, .
\label{equ:tworange}
\end{eqnarray}
Watch for the special case $(M_\text{eff})_{ij}=0$. The intersection of these
intervals is then obtained as
\begin{equation}
\underbrace{\max \left( \frac{(M_\text{eff}^{\mathrm{exp}})_{ij}^\mathrm{L}}{(M_\text{eff}^\text{th})_{ij}} \right)}_{c^L} \le   c  \le \underbrace{ \min \left( \frac{(M_\text{eff}^{\mathrm{exp}})_{ij}^\mathrm{U}}{(M_\text{eff}^\text{th})_{ij}} \right)}_{c^U}
\label{equ:lowup}
\end{equation}
\end{subequations}
with $(M_\text{eff}^{\mathrm{exp}})_{ij}^\mathrm{L}$ and $(M_\text{eff}^{\mathrm{exp}})_{ij}^\mathrm{U}$ chosen according to the case selection in \equ{tworange} for $i$ and $j$ individually. Only, if $c^L \le c^U$, then we have an allowed range for $c$.
If, however, $c^L > c^U$, the realization is refused. Note that the constant $c$ is not used
anymore further on since it may have different origins. However, the appropriate absolute mass scale will be implicitly determined in order to satisfy \equ{matching}. 

\subsection{Complexity and Counting of Cases}
\label{sec:counting}

In order to avoid in our procedure the generation of equivalent
realizations and double-counting of cases, several heuristics can be
used. As far as the phases in \equ{meffth} are concerned, we only have the matrices $D_D$, $\tilde{K}$,
$\tilde{D}$ for the real case, which leads to $2^{(3+2+3)}=2^8=256$ phases.
For $\theta_{13}^x=0$, the phase $\delta^x$ is unphysical, and we only test one case.
This leads to $4^6 \times (3 \times 2 + 1)^3$ different angle combinations in $M_\text{eff}^\text{th}$.
For $M_\text{eff}^\text{exp}$ in \equ{meffexp}, we do not generate $D_\ell$ in \equ{meffexp}, because the phases in the corresponding $D_D$ will
match a valid case of $M_\text{eff}^\text{exp}$ since both matrices appear on the outside 
(a valid phase in $D_D$ then  actually corresponds to two cases for $D_D$ and $D_\ell$).
In the real case, we therefore have only $2^2=4$ phases from $K_\ell$, and $4^2 \times (3 \times 2 +1)=112$
cases for the angles in $\widehat{U}_\ell$. This has to be multiplied with the
corresponding allowed hierarchies in $M_D$ and $M_R$. It is easy to derive that there are 
12 possible cases for a normal neutrino mass hierarchy, 9 possible cases for an inverted hierarchy,
and 13 cases for the degenerate case (\cf, Appendix \ref{sec:paramhier}). 
In addition, we test six cases for the true values in
$\widehat{U}_\text{PMNS}$.
Since in the degenerate case any $\widehat{U}_\ell$ and any $\widehat{U}_\text{PMNS}$ leads to a diagonal $M_\text{eff}^\text{exp}$, it is sufficient
to test one case for $\widehat{U}_\ell$ and $K_\ell$. This means that the degenerate case
hardly contributes to the complexity.
Further simplifications are possible for $M_D^\text{diag} = \mathbbm{1}$, \etc. In this case, 
one has to watch that the relative counting of different hierarchy/mixing angle cases is affected.
In total, we test
\begin{equation}
N = 256 \times 4^6 \times (3 \times 2 +1)^3 \times \left(112 \times 4 \times (12 + 9) +13 \right) \times 6 \simeq  20 \cdot 10^{12}
\end{equation}
different combinatorial possibilities, which require about 2 months of running time on a modern computer using a C-based software.

The number of allowed combinations may be used as a statistical measure in some cases. The interpretation
is then related to the number of different possibilities.

\section{Details of the Results for the NH Case}
\label{sec:detailsnh}

This section contains supplementary material for
\Sec~\ref{sec:nhtextures}, such as the selection criteria used and the
phases necessary for a complete reconstruction of the charged lepton
and neutrino Yukawa couplings.

\subsection{Texture Set Selection Criteria}
\label{sec:texselection}

Let us now describe the selection criteria leading to the
texture sets presented in \Sec~\ref{sec:nhtextures}. In producing
\Tab~\ref{tab:seesawtextures}, we have used the following criteria:
\begin{enumerate}
 \item  The seesaw realizations should resist an increased experimental pressure provided that the
current best-fit values are unchanged, \ie, $\stheta$ will not be
found. We impose an extrapolated experimental limit by using
$\sigma_{12}\simeq 4.6\%$~\cite{Minakata:2004jt} and
$\sigma_{23}\simeq 10\%$~\cite{Antusch:2004yx} in \equ{chi2}, which
corresponds to about a decade from now. Since $\theta_{13}$ is in any case smaller than $1^\circ$ and would anyway resist an increased experimental pressure, we do not have to consider it for a further selection.
\item Textures that differ by entries of the order $\epsilon^2$ should
  be included with only one example. We choose the realization with the best $\chi^2$. 
\item The selection of realizations should be stable under
  $\epsilon^2$-variations of the
mixing angles, \ie, if we find mixing angles $\epsilon^2$, the realization has to be valid for $\epsilon^2 \rightarrow 0$ as well. We include the newly obtained texture sets in \Tab~\ref{tab:seesawtextures} by
introducing the angle $\xi \in \{0, \epsilon^2 \}$. Note that these cases are initially
generated as well, but they might be filtered out later by \equ{matching}.

\item We omit texture sets with anarchical $M_R$ (matrices just filled
  with entries ``1''), since such structure-less textures do, in
  general, not yield much useful information.
\item We also show only one example for each texture set $(M_1,M_2,M_3)\sim(M_\ell,M_D,M_R)$, where $M_i$
  ($i=1,2,3$) are the corresponding textures obtained after texture
  reduction, which have one pair $\{M_j,M_k\}$ appearing together with more than one
  possible $M_i$ ($i \neq j \neq k$). In this case, we keep the
  texture set associated with a representation that has the lowest $\chi^2$.
\end{enumerate}
By using the above selection criteria, we obtain the list of 72 texture
sets shown in \Tabs~\ref{tab:seesawtextures}
and~\ref{tab:seesawtexturesphases}. The effects of the different selection criteria on the number of texture sets is shown in \Tab~\ref{tab:selectionnh}. In this table, we also
show the numbers of distinct textures for $M_\ell$, $M_D$ and $M_R$.
\begin{table}
\begin{center}
\begin{tabular}{c||c|ccc}
\hline
Selection criterion & \#Textures & $\#M_\ell$ & $\#M_D$ & $\#M_R$ \\
\hline 
 None & 1 981 & 20 & 621 & 35 \\ 
1.--2. & 1 048 & 20 & 475 & 31 \\  
1.--3. & 447 & 20 & 270 & 22 \\ 
1.--5. & 72 & 17 & 65 & 21 \\
\hline
\end{tabular}
\end{center}
\mycaption{\label{tab:selectionnh} Total number of texture sets, and the numbers
  of distinct textures for $M_\ell$, $M_D$ and $M_R$. Each row
  corresponds to the application of the selection criteria specified in the first column (see text).}
\end{table}

\subsection{Supplementary Information for the Texture Sets}
\label{sec:realizationsrest}

\Tab~\ref{tab:seesawtexturesphases} shows extra details of the realizations listed in \Tab~\ref{tab:seesawtextures}. It contains the complete set of phases, the PMNS mixing angles, $\chi^2$ (in the 10 years limit, \cf, \App~\ref{sec:texselection}), and the number of realizations leading to each texture set (for $\xi=\epsilon^2$ in ambiguous cases). Note that we chose $\varphi^{D'}_1=\varphi^{D'}_2=\varphi^{D'}_3=\alpha^{D}_1=\alpha^{D}_2=0$ since these phases appear in $M_\text{eff}^\text{th}$ only in combination with other phases (see \equ{mthmexpeff}), and can be absorbed into the other phases. \Tabs~\ref{tab:seesawtextures} and~\ref{tab:seesawtexturesphases} provide together the complete information sufficient to fully reconstruct the Yukawa coupling matrices of the 72 realizations.

\begin{small}
\begin{center}
\begin{longtable}{|@{\hspace{0.5mm}}c@{\hspace{0.5mm}}||@{\hspace{0.5mm}}c@{\hspace{0.5mm}}|@{\hspace{0.5mm}}c@{\hspace{0.5mm}}|@{\hspace{0.5mm}}c@{\hspace{0.5mm}}|@{\hspace{0.5mm}}c@{\hspace{0.5mm}}||@{\hspace{0.5mm}}c@{\hspace{0.5mm}}|@{\hspace{0.5mm}}c@{\hspace{0.5mm}}||@{\hspace{0.5mm}}c@{\hspace{0.5mm}}|}

\hline &&&&&&& \vspace{-3mm}
\\*
\#
& $ ( \delta ^l \hspace{-0.5mm}, \alpha _1^l , \alpha _2^l ) $ &
$( \delta ^D \hspace{-0.5mm}, \varphi _1^D , \varphi _2^D , \varphi _3^ D ) $ &
$( \delta ^{D'} \hspace{-1.5mm}, \alpha _1^{D'} , \alpha _2^{D'} ) $ &
$( \delta ^R \hspace{-0.5mm}, \varphi _1^R , \varphi _2^R , \varphi _3^R ) $ &
$( \theta_{12} , \theta_{13} , \theta_{23} ) $ &
$\chi^2$ &
Cases
\\*[1mm]\hline

$1$ & $

\begin{array}{c}
(0 , \pi , 0)
\end{array}
$ & $
\begin{array}{c}
(\pi ,0,0,\pi )
\end{array}
$ & $
\begin{array}{c}
(\pi ,\pi ,\pi )
\end{array}
$ & $
\begin{array}{c}
(0,0,0,\pi )
\end{array}
$ & $
\begin{array}{c}
(34.0^\circ,0.2^\circ,52.2^\circ)
\end{array}
$ & $7.12$ & 18
\\

$2$ & $

\begin{array}{c}
(\pi ,0,\pi )
\end{array}
$ & $
\begin{array}{c}
(\pi ,0,0,\pi )
\end{array}
$ & $
\begin{array}{c}
(0,0,\pi )
\end{array}
$ & $
\begin{array}{c}
(\pi ,0,\pi ,\pi )
\end{array}
$ & $
\begin{array}{c}
(33.6^\circ,0.2^\circ,51.5^\circ)
\end{array}
$ & $5.29$ & 38

\\

$3$ & $

\begin{array}{c}
(\pi ,0,0)
\end{array}
$ & $
\begin{array}{c}
(0,0,\pi ,0)
\end{array}
$ & $
\begin{array}{c}
(0,0,0)
\end{array}
$ & $
\begin{array}{c}
(\pi ,0,\pi ,0)
\end{array}
$ & $
\begin{array}{c}
(33.5^\circ,0.2^\circ,51.3^\circ)
\end{array}
$ & $4.9$ & 26

\\

$4$ & $

\begin{array}{c}
(0,0,\pi )
\end{array}
$ & $
\begin{array}{c}
(0,0,0,\pi )
\end{array}
$ & $
\begin{array}{c}
(\pi ,0,\pi )
\end{array}
$ & $
\begin{array}{c}
(\pi ,0,\pi ,\pi )
\end{array}
$ & $
\begin{array}{c}
(33.5^\circ,0.1^\circ,51.2^\circ)
\end{array}
$ & $4.71$ & 17

\\
\end{longtable}
\end{center}
\end{small}


\begin{thebibliography}{10}
\expandafter\ifx\csname bibnamefont\endcsname\relax
  \def\bibnamefont#1{#1}\fi
\expandafter\ifx\csname bibfnamefont\endcsname\relax
  \def\bibfnamefont#1{#1}\fi
\expandafter\ifx\csname url\endcsname\relax
  \def\url#1{\texttt{#1}}\fi
\expandafter\ifx\csname urlprefix\endcsname\relax\def\urlprefix{URL }\fi
\providecommand{\bibinfo}[2]{#2}
\providecommand{\eprint}[2][]{\url{#2}}

\bibitem{Fukuda:2002pe}
\bibinfo{author}{\bibfnamefont{S.}~\bibnamefont{Fukuda}} \emph{et~al.}
  (\bibinfo{collaboration}{Super-Kamiokande}), \bibinfo{journal}{Phys. Lett.}
  \textbf{\bibinfo{volume}{B539}}, \bibinfo{pages}{179} (\bibinfo{year}{2002}),
  \eprint{hep-ex/0205075}.

\bibitem{Ahmad:2002ka}
\bibinfo{author}{\bibfnamefont{Q.~R.} \bibnamefont{Ahmad}} \emph{et~al.}
  (\bibinfo{collaboration}{SNO}), \bibinfo{journal}{Phys. Rev. Lett.}
  \textbf{\bibinfo{volume}{89}}, \bibinfo{pages}{011302}
  (\bibinfo{year}{2002}), \eprint[http://arXiv.org/abs]{nucl-ex/0204009}.

\bibitem{Fukuda:1998mi}
\bibinfo{author}{\bibfnamefont{Y.}~\bibnamefont{Fukuda}} \emph{et~al.}
  (\bibinfo{collaboration}{Super-Kamiokande}), \bibinfo{journal}{Phys. Rev.
  Lett.} \textbf{\bibinfo{volume}{81}}, \bibinfo{pages}{1562}
  (\bibinfo{year}{1998}), \eprint{hep-ex/9807003}.

\bibitem{Araki:2004mb}
\bibinfo{author}{\bibfnamefont{T.}~\bibnamefont{Araki}} \emph{et~al.}
  (\bibinfo{collaboration}{KamLAND}), \bibinfo{journal}{Phys. Rev. Lett.}
  \textbf{\bibinfo{volume}{94}}, \bibinfo{pages}{081801}
  (\bibinfo{year}{2005}), \eprint{hep-ex/0406035}.

\bibitem{Apollonio:2002gd}
\bibinfo{author}{\bibfnamefont{M.}~\bibnamefont{Apollonio}} \emph{et~al.}
  (\bibinfo{collaboration}{CHOOZ}), \bibinfo{journal}{Eur. Phys. J.}
  \textbf{\bibinfo{volume}{C27}}, \bibinfo{pages}{331} (\bibinfo{year}{2003}),
  \eprint{hep-ex/0301017}.

\bibitem{Aliu:2004sq}
\bibinfo{author}{\bibfnamefont{E.}~\bibnamefont{Aliu}} \emph{et~al.}
  (\bibinfo{collaboration}{K2K}), \bibinfo{journal}{Phys. Rev. Lett.}
  \textbf{\bibinfo{volume}{94}}, \bibinfo{pages}{081802}
  (\bibinfo{year}{2005}), \eprint{hep-ex/0411038}.

\bibitem{SU5}
\bibinfo{journal}{H. Georgi and S. L. Glashow, Phys. Rev. Lett. {\bf 32}, 438
  (1974); H. Georgi, in {\it Proceedings of Coral Gables 1975, Theories and
  Experiments in High Energy Physics}, New York, 1975} .

\bibitem{PatiSalam}
\bibinfo{journal}{J. C. Pati and A. Salam, Phys. Rev. {\bf D8}, 1240 (1973);
  {\it ibid.} {\bf D10}, 275 (1974)} .

\bibitem{typeIseesaw}
\bibinfo{journal}{P. Minkowski, Phys. Lett. {\bf B67}, 421 (1977); T. Yanagida,
  in {\it Proceedings of the Workshop on the Unified Theory and Baryon Number
  in the Universe}, KEK, Tsukuba, 1979; M. Gell-Mann, P. Ramond, and R.
  Slansky, in {\it Proceedings of the Workshop on Supergravity}, North-Holland, Amsterdam, 1980; S. L. Glashow, in {\it Proceedings of the 1979 Cargese Summer
  Institute on Quarks and Leptons}, Plenum Press, New York, 1980} .

\bibitem{typeIIseesaw}
\bibinfo{journal}{M. Magg and C. Wetterich, Phys. Lett. {\bf B94}, 61 (1980);
  R. N. Mohapatra and G. Senjanovi\'c, Phys. Rev. Lett. {\bf 44}, 912 (1980);
  Phys. Rev. {\bf D23}, 165 (1981); J. Schechter and J.W.F. Valle, Phys. Rev.
  {\bf D22}, 2227 (1980); G. Lazarides, Q. Shafi, and C. Wetterich, Nucl. Phys.
  {B181}, 287 (1981)} .

\bibitem{GUTscale}
\bibinfo{journal}{H. Georgi and H. Quinn, Phys. Rev. Lett. {\bf 33}, 451
  (1974); S. Dimopoulos, S. Raby, and F. Wilczek, Phys. Rev. {\bf D24}, 1681
  (1981); S. Dimopoulos and H. Georgi, Nucl. Phys. {\bf B193}, 150 (1981)} .

\bibitem{CKM}
\bibinfo{journal}{N. Cabibbo, Phys. Rev. Lett. {\bf 10}, 531 (1963); M.
  Kobayashi and T. Maskawa, Prog. Theor. Phys. {\bf 49}, 652 (1973)} .

\bibitem{PMNS}
\bibinfo{journal}{B. Pontecorvo, Sov. Phys. JETP {\bf 6}, 429 (1957); Z. Maki,
  M. Nakagawa, and S. Sakata, Prog. Theor. Phys. {\bf 28}, 870 (1962)} .

\bibitem{Schwetz:2006dh}
\bibinfo{author}{\bibfnamefont{T.}~\bibnamefont{Schwetz}},
  \bibinfo{journal}{Phys. Scripta} \textbf{\bibinfo{volume}{T127}},
  \bibinfo{pages}{1} (\bibinfo{year}{2006}), \eprint{hep-ph/0606060}.

\bibitem{qlc}
\bibinfo{journal}{A. Y. Smirnov, {\tt hep-ph/0402264}; M. Raidal, Phys. Rev.
  Lett. {\bf 93}, 161801 (2004), {\tt hep-ph/0404046}; H. Minakata and A. Y.
  Smirnov, Phys. Rev. {\bf D70}, 073009 (2004), {\tt hep-ph/0405088}} .

\bibitem{Petcov:1993rk}
\bibinfo{author}{\bibfnamefont{S.~T.} \bibnamefont{Petcov}} \bibnamefont{and}
  \bibinfo{author}{\bibfnamefont{A.~Y.} \bibnamefont{Smirnov}},
  \bibinfo{journal}{Phys. Lett.} \textbf{\bibinfo{volume}{B322}},
  \bibinfo{pages}{109} (\bibinfo{year}{1994}), \eprint{hep-ph/9311204}.

\bibitem{tribimaximal}
\bibinfo{journal}{P.F. Harrison, D.H. Perkins, and W.G. Scott, Phys. Lett. {\bf
  B458}, 79 (1999), {\tt hep-ph/9904297}; Phys. Lett. {\bf B530}, 167 (2002),
  {\tt hep-ph/0202074}} .

\bibitem{qlcbimax}
\bibinfo{journal}{M. Jezabek and Y. Sumino, Phys. Lett. {\bf B457}, 139
  (1999),{\tt hep-ph/9904382}; C. Giunti and M. Tanimoto, Phys. Rev. {\bf D66},
  113006 (2002), {\tt hep-ph/0209169}; P. H. Frampton, S. T. Petcov, and W.
  Rodejohann, Nucl. Phys. {\bf B687}, 31 (2004), {\tt hep-ph/0401206}} .

\bibitem{qlcsumrules}
\bibinfo{journal}{T. Ohlsson, Phys. Lett. {\bf B622}, 159 (2005), {\tt
  hep-ph/0506094}; S. Antusch and S. F. King, Phys. Lett. {\bf B631}, 42
  (2005), {\tt hep-ph/0508044}} .

\bibitem{qlcpheno}
\bibinfo{journal}{K. Cheung, S. K. Kang, C. S. Kim, and J. Lee, Phys. Rev. {\bf
  D72}, 036003 (2005), {\tt hep-ph/0503122}; K. A. Hochmuth and W. Rodejohann,
  Phys. Rev. {\bf D75}, 073001 (2007),{\tt hep-ph/0607103}} .

\bibitem{qlcCabibbo}
\bibinfo{journal}{W. Rodejohann, Phys. Rev. {\tt D69}, 033005 (2004), {\tt
  hep-ph/0309249}; N. Li and B.-Q. Ma, Phys. Rev. {\tt D71}, 097301 (2005),
  {\tt hep-ph/0501226}; Z.-z. Xing, Phys. Lett., {\bf B618}, 141 (2005), {\tt
  hep-ph/0503200}; A. Datta, L. L. Everett, and P. Ramond, Phys. Lett. {\bf B620},
  42 (2005), {\tt hep-ph/0503222}; L. L. Everett, Phys. Rev. {\bf D73}, 013011
  (2006), {\tt hep-ph/0510256}} .

\bibitem{Chauhan:2006im}
\bibinfo{author}{\bibfnamefont{B.~C.} \bibnamefont{Chauhan}},
  \bibinfo{author}{\bibfnamefont{M.}~\bibnamefont{Picariello}},
  \bibinfo{author}{\bibfnamefont{J.}~\bibnamefont{Pulido}}, \bibnamefont{and}
  \bibinfo{author}{\bibfnamefont{E.}~\bibnamefont{Torrente-Lujan}},
  \bibinfo{journal}{Eur. Phys. J.} \textbf{\bibinfo{volume}{C50}},
  \bibinfo{pages}{573} (\bibinfo{year}{2007}), \eprint{hep-ph/0605032}.

\bibitem{qlcRG}
\bibinfo{journal}{A. Dighe, S. Goswami, and P. Roy, Phys. Rev. {\bf D73},
  071301 (2006), {\tt hep-ph/0602062}; M. A. Schmidt and A. Y. Smirnov, {\tt
  hep-ph/0607232}} .

\bibitem{qlcmodels}
\bibinfo{journal}{T. Ohlsson and G. Seidl, Nucl. Phys. {\bf B643}, 247 (2002),
  {\tt hep-ph/0206087}; P. H. Frampton and R. N. Mohapatra, JHEP {\bf 01}, 025
  (2005), {\tt hep-ph/0407139}; S. Antusch, S. F. King, and R. N.
  Mohapatra, Phys. Lett.{\bf B618}, 150 (2005), {\tt hep-ph/0504007}; M.
  Picariello, {\tt hep-ph/0611189}} .

\bibitem{Plentinger:2006nb}
\bibinfo{author}{\bibfnamefont{F.}~\bibnamefont{Plentinger}},
  \bibinfo{author}{\bibfnamefont{G.}~\bibnamefont{Seidl}}, \bibnamefont{and}
  \bibinfo{author}{\bibfnamefont{W.}~\bibnamefont{Winter}}
  (\bibinfo{year}{2006}), \eprint{hep-ph/0612169}.

\bibitem{Casas:2006hf}
\bibinfo{author}{\bibfnamefont{J.~A.} \bibnamefont{Casas}},
  \bibinfo{author}{\bibfnamefont{A.}~\bibnamefont{Ibarra}}, \bibnamefont{and}
  \bibinfo{author}{\bibfnamefont{F.}~\bibnamefont{Jimenez-Alburquerque}},
  \bibinfo{journal}{JHEP} \textbf{\bibinfo{volume}{04}}, \bibinfo{pages}{064}
  (\bibinfo{year}{2007}), \eprint{hep-ph/0612289}.

\bibitem{chargedleptons}
\bibinfo{journal}{G. Altarelli, F. Feruglio, and I. Masina, Nucl. Phys. {\bf
  B689}, 157 (2004), {\tt hep-ph/0402155}; A. Romanino, Phys. Rev. {\bf D70},
  013003 (2004), {\tt hep-ph/0402258}; S. Antusch and S. F. King, Phys. Lett.
  {\bf B591}, 104 (2004), {\tt hep-ph/0403053}; C. A. de S. Pires, J. Phys.
  {\bf G30}, B29 (2004), {\tt hep-ph/0404146}; K. A. Hochmuth, S. T. Petcov,
  and W. Rodejohann, {\tt arXiv:0706.2975 [hep-ph]}} .

\bibitem{Wolfenstein:1983yz}
\bibinfo{author}{\bibfnamefont{L.}~\bibnamefont{Wolfenstein}},
  \bibinfo{journal}{Phys. Rev. Lett.} \textbf{\bibinfo{volume}{51}},
  \bibinfo{pages}{1945} (\bibinfo{year}{1983}).

\bibitem{Blucher:2005dc}
\bibinfo{author}{\bibfnamefont{E.}~\bibnamefont{Blucher}} \emph{et~al.}
  (\bibinfo{year}{1100}), \eprint{hep-ph/0512039}.

\bibitem{GSTO}
\bibinfo{journal}{R. Gatto, G. Sartori, and M. Tonin, Phys. Lett. {\bf B28},
  128 (1968); R. J. Oakes, Phys. Lett. {\bf B29}, 683 (1969); Phys. Lett. {\bf
  B31}, 620 (E) (1970); Phys. Lett. {\bf B30}, 262 (1969)} .

\bibitem{Chankowski:2005qp}
\bibinfo{author}{\bibfnamefont{P.~H.} \bibnamefont{Chankowski}},
  \bibinfo{author}{\bibfnamefont{K.}~\bibnamefont{Kowalska}},
  \bibinfo{author}{\bibfnamefont{S.}~\bibnamefont{Lavignac}}, \bibnamefont{and}
  \bibinfo{author}{\bibfnamefont{S.}~\bibnamefont{Pokorski}},
  \bibinfo{journal}{Phys. Rev.} \textbf{\bibinfo{volume}{D71}},
  \bibinfo{pages}{055004} (\bibinfo{year}{2005}), \eprint{hep-ph/0501071}.

\bibitem{Gonzalez-Garcia:2004jd}
\bibinfo{author}{\bibfnamefont{M.~C.} \bibnamefont{Gonzalez-Garcia}},
  \bibinfo{journal}{Phys. Scripta} \textbf{\bibinfo{volume}{T121}},
  \bibinfo{pages}{72} (\bibinfo{year}{2005}), \eprint{hep-ph/0410030}.

\bibitem{Enkhbat:2005xb}
\bibinfo{author}{\bibfnamefont{T.}~\bibnamefont{Enkhbat}} \bibnamefont{and}
  \bibinfo{author}{\bibfnamefont{G.}~\bibnamefont{Seidl}},
  \bibinfo{journal}{Nucl. Phys.} \textbf{\bibinfo{volume}{B730}},
  \bibinfo{pages}{223} (\bibinfo{year}{2005}), \eprint{hep-ph/0504104}.

\bibitem{Grimus:2002zh}
\bibinfo{author}{\bibfnamefont{W.}~\bibnamefont{Grimus}} \bibnamefont{and}
  \bibinfo{author}{\bibfnamefont{L.}~\bibnamefont{Lavoura}},
  \bibinfo{journal}{Eur. Phys. J.} \textbf{\bibinfo{volume}{C28}},
  \bibinfo{pages}{123} (\bibinfo{year}{2003}), \eprint{hep-ph/0211334}.

\bibitem{earlyU1}
\bibinfo{journal}{J. Bijnens and C. Wetterich, Nucl. Phys. {\bf B283}, 237
  (1987); Phys. Lett. {\bf B199}, 525 (1987); M. Leurer, Y. Nir, and N.
  Seiberg, Nucl. Phys. {\bf B398}, 319 (1993), {\tt hep-ph/9212278}; Nucl.
  Phys. {\bf B420}, 468 (1994), {\tt hep-ph/9310320}; L. E. Ibanez and G. G.
  Ross, Phys. Lett. {\bf B332}, 100 (1994); P. Binetruy and P. Ramond, Phys.
  Lett. {\bf B350}, 49 (1995), {\tt hep-ph/9412385}; V. Jain and R. Shrock,
  Phys. Lett. {\bf B352}, 83 (1995), {\tt hep-ph/9412367}; E. Dudas, S.
  Pokorski, and C. A. Savoy, Phys. Lett. {\bf B356}, 45 (1995), {\tt
  hep-ph/9504292}; Y. Nir, Phys. Lett. {\bf B354}, 107 (1995), {\tt
  hep-ph/9504312}; P. Binetruy, S. Lavignac, and P. Ramond, Nucl. Phys. {\bf
  B477}, 353 (1996), {\tt hep-ph/9601243}} .

\bibitem{recentU1}
\bibinfo{journal}{Q. Shafi and Z. Tavartkiladze, Phys. Lett. {\bf B482}, 145
  (2000), {\tt hep-ph/0002150}; J. Ellis, G. K. Leontaris, and J. Rizos, JHEP
  {\bf 05}, 001 (2000), {\tt hep-ph/0002263}; A. J. Joshipura, R. D. Vaidya,
  and S. K. Vempati, Phys. Rev. {\bf D62}, 093020 (2000), {\tt hep-ph/0006138};
  N. Maekawa, Prog. Theor. Phys. {\bf 106}, 401 (2001), {\tt hep-ph/0104200};
  M. Kakizaki and M. Yamaguchi, JHEP {\bf 06}, 032 (2002), {\tt
  hep-ph/0203192}; K. S. Babu, T. Enkhbat, and I. Gogoladze, Nucl. Phys. {\bf
  B678}, 233 (2004), {\tt hep-ph/0308093}; I. Jack, D. R. T. Jones, and R.
  Wild, Phys. Lett. {\bf B580}, 72 (2004), {\tt hep-ph/0309165}; H. Dreiner, H.
  Murayama, and M. Thormeier, Nucl. Phys. {\bf B729}, 278 (2005), {\tt
  hep-ph/0312012}; Y. E. Antebi, Y. Nir, and T. Volansky, Phys. Rev. {\bf D73},
  075009 (2006), {\tt hep-ph/0512211}; H. K. Dreiner {\it et al.}, Nucl. Phys.
  {\bf B774}, 127 (2007), {\tt hep-ph/0610026}; J. R. Ellis, M. E. Gomez, and
  S. Lola (2006), {\tt hep-ph/0612292}; I. Gogoladze, C. A. Lee, T. J. Li, and
  Q. Shafi, {\tt arXiv:0705.3035 [hep-ph]}} .

\bibitem{discretequarks}
\bibinfo{journal}{E. Ma, Mod. Phys. Lett. {\bf A17}, 627 (2002), {\tt
  hep-ph/0203238}; G. Altarelli and F. Feruglio, Nucl. Phys. {\bf B741}, 215
  (2006), {\tt hep-ph/0512103}; S. F. King and M. Malinsky, JHEP {\bf 11}, 071
  (2006), {\tt hep-ph/0608021}; Phys. Lett. {\bf B645}, 351 (2007), {\tt
  hep-ph/0610250}; F. Feruglio, C. Hagedorn, Y. Lin, and L. Merlo, Nucl. Phys.
  {\bf B775}, 120 (2007), {\tt hep-ph/0702194}; C. Luhn, S. Nasri, and P.
  Ramond, {\tt arXiv:0706.2341 [hep-ph]}} .

\bibitem{discreteGUTs}
\bibinfo{journal}{E. Ma, H. Sawanaka, and M. Tanimoto, Phys. Lett. {\bf B641},
  301 (2006), {\tt hep-ph/0606103}; I. de Medeiros Varzielas, S. F. King, and
  G. G. Ross, Phys. Lett. {\bf B648}, 201 (2007), {\tt hep-ph/0607045}; E. Ma,
  Mod. Phys. Lett. {\bf A21}, 2931 (2006), {\tt hep-ph/0607190}; S. Morisi, M.
  Picariello, and E. Torrente-Lujan, Phys. Rev. {\bf D75}, 075015 (2007), {\tt
  hep-ph/0702034}; M.-C. Chen and K.T. Mahanthappa, {\tt arXiv:0705.0714
  [hep-ph]}} .

\bibitem{Altarelli:2007cd}
\bibinfo{author}{\bibfnamefont{G.}~\bibnamefont{Altarelli}}
  (\bibinfo{year}{0500}), \eprint{arXiv:0705.0860 [hep-ph]}.

\bibitem{Froggatt:1978nt}
\bibinfo{author}{\bibfnamefont{C.~D.} \bibnamefont{Froggatt}} \bibnamefont{and}
  \bibinfo{author}{\bibfnamefont{H.~B.} \bibnamefont{Nielsen}},
  \bibinfo{journal}{Nucl. Phys.} \textbf{\bibinfo{volume}{B147}},
  \bibinfo{pages}{277} (\bibinfo{year}{1979}).

\bibitem{rephasings}
\bibinfo{journal}{H. K. Dreiner {\it et al.}, {\tt hep-ph/0703074}; E. Jenkins
  and A. V. Manohar, {\tt arXiv:0706.4313 [hep-ph]}} .

\bibitem{Huber:2004ug}
\bibinfo{author}{\bibfnamefont{P.}~\bibnamefont{Huber}},
  \bibinfo{author}{\bibfnamefont{M.}~\bibnamefont{Lindner}},
  \bibinfo{author}{\bibfnamefont{M.}~\bibnamefont{Rolinec}},
  \bibinfo{author}{\bibfnamefont{T.}~\bibnamefont{Schwetz}}, \bibnamefont{and}
  \bibinfo{author}{\bibfnamefont{W.}~\bibnamefont{Winter}},
  \bibinfo{journal}{Phys. Rev.} \textbf{\bibinfo{volume}{D70}},
  \bibinfo{pages}{073014} (\bibinfo{year}{2004}), \eprint{hep-ph/0403068}.

\bibitem{Antusch:2004yx}
\bibinfo{author}{\bibfnamefont{S.}~\bibnamefont{Antusch}},
  \bibinfo{author}{\bibfnamefont{P.}~\bibnamefont{Huber}},
  \bibinfo{author}{\bibfnamefont{J.}~\bibnamefont{Kersten}},
  \bibinfo{author}{\bibfnamefont{T.}~\bibnamefont{Schwetz}}, \bibnamefont{and}
  \bibinfo{author}{\bibfnamefont{W.}~\bibnamefont{Winter}},
  \bibinfo{journal}{Phys. Rev.} \textbf{\bibinfo{volume}{D70}},
  \bibinfo{pages}{097302} (\bibinfo{year}{2004}), \eprint{hep-ph/0404268}.

\bibitem{Minakata:2004jt}
\bibinfo{author}{\bibfnamefont{H.}~\bibnamefont{Minakata}},
  \bibinfo{author}{\bibfnamefont{H.}~\bibnamefont{Nunokawa}},
  \bibinfo{author}{\bibfnamefont{W.~J.~C.} \bibnamefont{Teves}},
  \bibnamefont{and}
  \bibinfo{author}{\bibfnamefont{R.}~\bibnamefont{Zukanovich~Funchal}},
  \bibinfo{journal}{Phys. Rev.} \textbf{\bibinfo{volume}{D71}},
  \bibinfo{pages}{013005} (\bibinfo{year}{2005}), \eprint{hep-ph/0407326}.

\bibitem{Bandyopadhyay:2004cp}
\bibinfo{author}{\bibfnamefont{A.}~\bibnamefont{Bandyopadhyay}},
  \bibinfo{author}{\bibfnamefont{S.}~\bibnamefont{Choubey}},
  \bibinfo{author}{\bibfnamefont{S.}~\bibnamefont{Goswami}}, \bibnamefont{and}
  \bibinfo{author}{\bibfnamefont{S.~T.} \bibnamefont{Petcov}},
  \bibinfo{journal}{Phys. Rev.} \textbf{\bibinfo{volume}{D72}},
  \bibinfo{pages}{033013} (\bibinfo{year}{2005}), \eprint{hep-ph/0410283}.

\bibitem{Barger:2006kp}
\bibinfo{author}{\bibfnamefont{V.}~\bibnamefont{Barger}},
  \bibinfo{author}{\bibfnamefont{P.}~\bibnamefont{Huber}},
  \bibinfo{author}{\bibfnamefont{D.}~\bibnamefont{Marfatia}}, \bibnamefont{and}
  \bibinfo{author}{\bibfnamefont{W.}~\bibnamefont{Winter}}
  (\bibinfo{year}{2006}), \eprint{hep-ph/0610301}.

\bibitem{nufact}
\bibinfo{journal}{A. Cervera {\it et al.}, Nucl. Phys. {\bf B579}, 17 (2000),
  {\tt hep-ph/0002108}; P. Huber, M. Lindner, and W. Winter, Nucl. Phys. {\bf
  B645}, 3 (2002), {\tt hep-ph/0204352}; P. Huber, M. Lindner, M. Rolinec, and
  W. Winter, Phys. Rev. {\bf D74}, 073003 (2006), {\tt hep-ph/0606119}} .

\bibitem{Georgi:1979df}
\bibinfo{author}{\bibfnamefont{H.}~\bibnamefont{Georgi}} \bibnamefont{and}
  \bibinfo{author}{\bibfnamefont{C.}~\bibnamefont{Jarlskog}},
  \bibinfo{journal}{Phys. Lett.} \textbf{\bibinfo{volume}{B86}},
  \bibinfo{pages}{297} (\bibinfo{year}{1979}).

\bibitem{Arason:1991hu}
\bibinfo{author}{\bibfnamefont{H.}~\bibnamefont{Arason}} \emph{et~al.},
  \bibinfo{journal}{Phys. Rev. Lett.} \textbf{\bibinfo{volume}{67}},
  \bibinfo{pages}{2933} (\bibinfo{year}{1991}).

\bibitem{Arason:1992eb}
\bibinfo{author}{\bibfnamefont{H.}~\bibnamefont{Arason}},
  \bibinfo{author}{\bibfnamefont{D.~J.} \bibnamefont{Castano}},
  \bibinfo{author}{\bibfnamefont{E.~J.} \bibnamefont{Piard}}, \bibnamefont{and}
  \bibinfo{author}{\bibfnamefont{P.}~\bibnamefont{Ramond}},
  \bibinfo{journal}{Phys. Rev.} \textbf{\bibinfo{volume}{D47}},
  \bibinfo{pages}{232} (\bibinfo{year}{1993}), \eprint{hep-ph/9204225}.

\bibitem{neutrinoRGs}
\bibinfo{journal}{K. S. Babu, C. N. Leung, and J. T. Pantaleone, Phys. Lett.
  {\bf B319}, 191 (1993), {\tt hep-ph/9309223}; J. A. Casas {\it et al.}, Nucl.
  Phys. {\bf B569}, 82 (2000), {\tt hep-ph/9905381}; K. R. S. Balaji, A. S. Dighe, R. N. Mohapatra, and M. K. Parida, Phys. Rev. Lett. {\bf 84}, 5034 (2000), {\tt hep-ph/0001310}; S.
  Antusch {\it et al.}, Phys. Lett. {\bf B519}, 238 (2001), {\tt
  hep-ph/0108005}; Phys. Lett. {\bf B525}, 130 (2002), {\tt hep-ph/0110366}; P.
  H. Chankowski and S. Pokorski, Int. J. Mod. Phys. {\bf A17}, 575 (2002); S.
  Antusch {\it et al.}, Nucl. Phys. {\bf B674}, 401 (2003), {\tt
  hep-ph/0305273}} .

\bibitem{Dighe:2007ks}
\bibinfo{author}{\bibfnamefont{A.}~\bibnamefont{Dighe}},
  \bibinfo{author}{\bibfnamefont{S.}~\bibnamefont{Goswami}}, \bibnamefont{and}
  \bibinfo{author}{\bibfnamefont{P.}~\bibnamefont{Roy}}
  (\bibinfo{year}{2007}), \eprint{arXiv:0704.3735 [hep-ph]}.

\bibitem{Ellis:2005dr}
\bibinfo{author}{\bibfnamefont{J.~R.} \bibnamefont{Ellis}},
  \bibinfo{author}{\bibfnamefont{A.}~\bibnamefont{Hektor}},
  \bibinfo{author}{\bibfnamefont{M.}~\bibnamefont{Kadastik}},
  \bibinfo{author}{\bibfnamefont{K.}~\bibnamefont{Kannike}}, \bibnamefont{and}
  \bibinfo{author}{\bibfnamefont{M.}~\bibnamefont{Raidal}},
  \bibinfo{journal}{Phys. Lett.} \textbf{\bibinfo{volume}{B631}},
  \bibinfo{pages}{32} (\bibinfo{year}{2005}), \eprint{hep-ph/0506122}.

\bibitem{King:2003jb}
\bibinfo{author}{\bibfnamefont{S.~F.} \bibnamefont{King}},
  \bibinfo{journal}{Rept. Prog. Phys.} \textbf{\bibinfo{volume}{67}},
  \bibinfo{pages}{107} (\bibinfo{year}{2004}), \eprint{hep-ph/0310204}.

\bibitem{leptogenesis}
\bibinfo{journal}{L. Covi, E. Roulet, and F. Vissani, Phys. Lett. {\bf B384},
  169 (1996), {\tt hep-ph/9605319}; M. Plumacher, Nucl. Phys. {\bf B530}, 207
  (1998), {\tt hep-ph/9704231}; A. Pilaftsis, Int. J. Mod. Phys. {\bf A14},
  1811 (1999), {\tt hep-ph/9812256}; J. R. Ellis, S. and Lola, and D. V.
  Nanopoulos, Phys. Lett. {\bf B452}, 87 (1999), {\tt hep-ph/9902364}; W.
  Buchmuller and M. Plumacher, Int. J. Mod. Phys. {\bf A15}, 5047 (2000), {\tt
  hep-ph/0007176}; G. C. Branco, M. N. Rebelo, and J. Silva-Marcos, J., Phys.
  Lett. {\bf B633}, 345 (2006), {\tt hep-ph/0510412}} .

\bibitem{resleptogenesis}
\bibinfo{journal}{M. Flanz, E. A. Paschos, U. Sarkar, and J. Weiss, Phys. Lett.
  {\bf B389}, 693-699 (1996), {\tt hep-ph/9607310}; A. Pilaftsis and T. E. J.
  Underwood, Nucl. Phys. {\bf B692}, 303-345 (2004), {\tt hep-ph/0309342}; A.
  Anisimov, A. Broncano, and M. Plumacher, Nucl. Phys. {\bf 737}, 176-189
  (2006), {\tt hep-ph/0511248}} .

\bibitem{resonantmodels}
\bibinfo{journal}{R. Gonzalez Felipe, F. R. Joaquim, and B. M. Nobre, Phys.
  Rev. {\bf D70}, 085009 (2004), {\tt hep-ph/0311029}; T. Hambye, J.
  March-Russell, and S. M. West, JHEP {\bf 07}, 070 (2004), {\tt
  hep-ph/0403183}; G. C. Branco, R. Gonzalez Felipe, F. R. Joaquim, and B. M.
  Nobre, Phys. Lett. {\bf B633}, 336-344 (2006), {\tt hep-ph/0507092}; S. M.
  West, Mod. Phys. Lett. {\bf A21}, 1629 (2006); T. Gherghetta, K. Kadota, and
  M. Yamaguchi, {\tt arXiv:0705.1749 [hep-ph]}; K. S. Babu, A. G. Bachri, and
  Z. Tavartkiladze, {\tt arXiv:0705.4419 [hep-ph]}} .

\bibitem{flavorleptogenesis}
\bibinfo{journal}{A. Abada \textit{et al.}, JHEP {\bf 09}, 010 (2006), {\tt
  hep-ph/0605281}; G. C. Branco, R. Gonzalez Felipe, and F. R. Joaquim, Phys.
  Lett. {\bf B645}, 432-436 (2007), {\tt hep-ph/0609297}} .

\bibitem{Pascoli:2006ie}
\bibinfo{author}{\bibfnamefont{S.}~\bibnamefont{Pascoli}},
  \bibinfo{author}{\bibfnamefont{S.~T.} \bibnamefont{Petcov}},
  \bibnamefont{and} \bibinfo{author}{\bibfnamefont{A.}~\bibnamefont{Riotto}},
  \bibinfo{journal}{Phys. Rev.} \textbf{\bibinfo{volume}{D75}},
  \bibinfo{pages}{083511} (\bibinfo{year}{2007}), \eprint{hep-ph/0609125}.

\bibitem{TexWeb}
\bibinfo{author}{\bibfnamefont{F.}~\bibnamefont{Plentinger}},
  \bibinfo{author}{\bibfnamefont{G.}~\bibnamefont{Seidl}}, \bibnamefont{and}
  \bibinfo{author}{\bibfnamefont{W.}~\bibnamefont{Winter}},
  \emph{\bibinfo{title}{Seesaw texture web page}} (\bibinfo{year}{2007}),
  \bibinfo{note}{{\tt
  http://theorie.physik.uni-wuerzburg.de/$\sim$winter/Resources/SeeSawTex}}.

\bibitem{Batra:2005rh}
\bibinfo{author}{\bibfnamefont{P.}~\bibnamefont{Batra}},
  \bibinfo{author}{\bibfnamefont{B.~A.} \bibnamefont{Dobrescu}},
  \bibnamefont{and} \bibinfo{author}{\bibfnamefont{D.}~\bibnamefont{Spivak}},
  \bibinfo{journal}{J. Math. Phys.} \textbf{\bibinfo{volume}{47}},
  \bibinfo{pages}{082301} (\bibinfo{year}{2006}), \eprint{hep-ph/0510181}.

\bibitem{anarchy}
\bibinfo{journal}{T. Goldman and G. J. Stephenson, Phys. Rev. {\bf D24},
  236 (1981); L. J. Hall, H. Murayama, and N. Weiner, Phys. Rev. Lett., {\bf
  84}, 2572 (2000), {\tt hep-ph/9911341}; N. Haba and H. Murayama, Phys. Rev.
  {\bf D63}, 053010 (2001), {\tt hep-ph/0009174}; G. Altarelli, F. Feruglio,
  and I. Masina, JHEP {\bf 01}, 035 (2003), {\tt hep-ph/0210342}; A. de Gouvea
  and H. Murayama, Phys. Lett. {\bf B573}, 94 (2003), {\tt hep-ph/0301050}; J.
  R. Espinosa, {\tt hep-ph/0306019}} .

\bibitem{statmasses}
\bibinfo{journal}{J. F. Donoghue, K. Dutta, and A. Ross, Phys. Rev. {\bf D73},
  113002 (2006), {\tt hep-ph/0511219}; B. Feldstein, L. J. Hall, and T. Watari,
  Phys. Rev. {\bf D74}, 095011 (2006), {\tt hep-ph/0608121}} .

\end{thebibliography}

\end{document}